\documentclass[
floatfix,
 aip,
 amsmath,amssymb,
 reprint,%
]{revtex4-2}

\usepackage{graphicx}
\usepackage{dcolumn}
\usepackage{bm}
\pdfoutput=1

\usepackage[utf8]{inputenc}
\usepackage[T1]{fontenc}
\usepackage{mathptmx}
\usepackage{etoolbox}
\usepackage{xspace}
\usepackage{color}
\usepackage{braket}
\usepackage[]{xcolor}

\makeatletter
\def\@email#1#2{%
 \endgroup
 \patchcmd{\titleblock@produce}
  {\frontmatter@RRAPformat}
  {\frontmatter@RRAPformat{\produce@RRAP{*#1\href{mailto:#2}{#2}}}\frontmatter@RRAPformat}
  {}{}
}%
\makeatother
\begin{document}

\newcommand{\Csixty}{\ensuremath{\mathrm{C_{60}}}\xspace}
\newcommand{\Cseventy}{\ensuremath{\mathrm{C_{70}}}\xspace}
\newcommand{\Htwo}{\ensuremath{\mathrm{H_2}}\xspace}
\newcommand{\HeCsixty}[1]{\ensuremath{\mathrm{^{#1}He}}@\ensuremath{\mathrm{C_{60}}}\xspace}
\newcommand{\HtwoCsixty}{\ensuremath{\mathrm{H_2}}@\ensuremath{\mathrm{C_{60}}}\xspace}
\newcommand{\HtwoOCsixty}{\ensuremath{\mathrm{H_2O}}@\ensuremath{\mathrm{C_{60}}}\xspace}
\newcommand{\HFCsixty}{\ensuremath{\mathrm{HF}}@\ensuremath{\mathrm{C_{60}}}\xspace}
\newcommand{\MethaneCsixty}{\ensuremath{\mathrm{CH_4}}@\ensuremath{\mathrm{C_{60}}}\xspace}
\newcommand{\HtwoCseventy}{\ensuremath{\mathrm{H_2}}@\ensuremath{\mathrm{C_{70}}}\xspace}
\newcommand{\Helium}[1]{\ensuremath{{}^{#1}\mathrm{He}}\xspace}

\newcommand{\Cth}{\ensuremath{{}^{13}\mbox{C}}\xspace}
\newcommand{\Cthtwo}{\ensuremath{{}^{13}\mbox{C}_2}\xspace}
\newcommand{\Ctw}{\ensuremath{{}^{12}\mbox{C}}\xspace}

\newcommand{\ODCBd}[1]{\ensuremath{\mbox{ODCB-d}_#1}\xspace}

\newcommand{\vecr}{\ensuremath{\mathbf{r}}}
\newcommand{\vecq}{\ensuremath{\mathbf{q}}}
\newcommand{\SpherHar}[2]{\ensuremath{Y^{#1}_{#2}}}
\def\ket#1{\left| #1 \right>}
\def\bra#1{\left< #1 \right|}
\newcommand{\ml}{{m}}

\newcommand{\angularEQ}{\sin(\theta)\frac{\partial}{\partial \theta}\left(\sin(\theta)\frac{\partial Y_\vecq}{\partial \theta}\right) + \frac{\partial^2 Y_\vecq}{\partial \phi^2} & = -\ell(\ell+1)\sin^2(\theta)Y_\vecq}

\newcommand{\radialEQ}{-\frac{\hbar}{2M}\frac{\partial^2 u_\vecq}{\partial r^2}+\left[V(r)+\frac{\hbar^2}{2M}\frac{\ell(\ell+1)}{r^2}\right]u_\vecq & =E_\vecq u_\vecq}

\newcommand{\Enl}{E_{n \ell}}

\newcommand{\PSInlm}{\psi_{n \ell m}}

\newcommand{\Rnl}{R_{n\ell}}


\newcommand{\Ylm}{Y_{\ell m}}

\newcommand{\YsphericalHarmonics}{\Ylm (\theta,\phi) & =\sqrt{\frac{(2\ell+1)}{4\pi}\frac{(\ell-m)!}{(\ell+m)!}}e^{i m\phi}P_\ell^{m}(cos(\theta))}

\newcommand{\PLegendreAssocFct}{P_\ell^m(x)&=(-1)^m (1-x^2)^{m/2} (\frac{d}{dx})^m P_\ell(x)}

\newcommand{\PLegendrePoly}{P_\ell(x)&=\frac{1}{2^\ell \ell!}(\frac{d}{dx})^\ell(x^2-1)^\ell}


\newcommand{\RwfHOn}{\Rnl(r)&=\frac{N_{\ell}^{n}}{b^{3/2}}(\frac{r}{b})^\ell e^{-r^2/2b^2}L_{n}^{(\ell+\frac{1}{2})}(\frac{r^2}{b^2})}

\newcommand{\RwfHOAnormWn}{N_{\ell}^{n}& = \sqrt{\frac{2^{n_r+\ell+2}n_r!}{\sqrt{\pi}(2 n_r + 2 \ell + 1)!!}}}

\newcommand{\RwfHO}{\Rnl(r)&=\frac{N_{\ell}^{\orange{n_r}}}{b^{3/2}}(\frac{r}{b})^\ell e^{-r^2/2b^2}L_{\orange{n_r}}^{(\ell+\frac{1}{2})}(\frac{r^2}{b^2})}

\newcommand{\RwfHOAnorm}{N_{\ell}^{\orange{n_r}}& = \sqrt{\frac{2^{n_r+\ell+2}n_r!}{\sqrt{\pi}(2 n_r + 2 \ell + 1)!!}}}


\newcommand{\matrixELEMENTgeneral}{\braket{\Psi_i|\hat{H}^{(a)}|\Psi_j} =\int_0^\infty\int_0^\pi\int_0^{2\pi}\Psi_i^*\hat{H}^{(a)}\Psi_j \, r^2 \, \sin(\theta) \; dr \; d\theta \; d\phi}

\newcommand{\THzLineIntesity}{S(\omega_{if}) & =\frac{N \pi f}{V \eta c \epsilon_0 \hbar} p_i \, \omega_{if} |\braket{\Psi_f|\hat{d}|\Psi_i} |^2}

\newcommand{\THzdipMOM}{d^{1}_m & =a^{1}_m \, r \, \SpherHar{1}{m}(\theta,\phi)}

\newcommand{\He}[1]{\ensuremath{\mathrm{^{#1}He}}\xspace}
\newcommand{\Schrodinger}{Schr{\"o}dinger\xspace}
\newcommand{\Ih}{\ensuremath{\mathrm{I_h}}\xspace}
\newcommand{\Bacic}{Ba{\v c}i{\'c}\xspace}
\newcommand{\blue}[1]{\textcolor{blue}{#1}}
\newcommand{\green}[1]{\textcolor{ForestGreen}{#1}}
\newcommand{\orange}[1]{\textcolor{orange}{#1}}
\newcommand{\red}[1]{\textcolor{red}{#1}}
\newcommand{\MHLnote}[1]{\blue{[MHL: #1]}}
\newcommand{\TRnote}[1]{\red{[TR: #1]}}
\newcommand{\MHLToHere}{\ \newline\MHLnote{** to here **}\newline\ }
\newcommand{\GRBnote}[1]{\orange{[GRB: #1]}}
\newcommand{\JRnote}[1]{\orange{[JR: #1]}}
\newcommand{\unit}[1]{\ensuremath{\mathrm{\ #1}}\xspace}
\newcommand{\picom}{\unit{pm}}
\newcommand{\wn}{\unit{{cm}^{-1}}}
\newcommand{\meV}{\unit{meV}}
\newcommand{\kel}{\unit{K}}
\newcommand{\T}{\unit{T}}
\newcommand{\MHz}{\unit{MHz}}
\newcommand{\meVperpmsq}{\unit{meV\,pm^{-2}}}
\newcommand{\meVperpmfo}{\unit{meV\,pm^{-4}}}
\newcommand{\meVperpmsx}{\unit{meV\,pm^{-6}}}

\newcommand{\angstr}{\unit{\AA}}

\newcommand{\meVperANGsq}{\unit{meV\,\text{\AA}^{-2}}}
\newcommand{\meVperANGfo}{\unit{meV\,\text{\AA}^{-4}}}
\newcommand{\meVperANGsx}{\unit{meV\,\text{\AA}^{-6}}}

\preprint{AIP/123-QED}


\title{Experimental Determination of the Interaction Potential between a Helium Atom and the Interior Surface of a $\mathrm{C_{60}}$ Fullerene Molecule}

\author{George Razvan Bacanu}
\affiliation{School of Chemistry, University of Southampton,  Southampton, SO17~1BJ, UK}

\author{Tanzeeha Jafari}
\affiliation{National Institute of Chemical Physics and Biophysics, Tallinn, 12618, Estonia}

\author{Mohamed  Aouane}
\affiliation{Institut Laue-Langevin, BP 156, 38042 Grenoble, France}

\author{Jyrki Rantaharju}
\affiliation{School of Chemistry, University of Southampton,  Southampton, SO17~1BJ, UK}

\author{Mark Walkey}
\affiliation{School of Chemistry, University of Southampton,  Southampton, SO17~1BJ, UK}

\author{Gabriela Hoffman}
\affiliation{School of Chemistry, University of Southampton,  Southampton, SO17~1BJ, UK}

\author{Anna Shugai}
\affiliation{National Institute of Chemical Physics and Biophysics, Tallinn, 12618, Estonia}

\author{Urmas Nagel}
\affiliation{National Institute of Chemical Physics and Biophysics, Tallinn, 12618, Estonia}

\author{Monica Jim{\'e}nez-Ruiz}
\affiliation{Institut Laue-Langevin, BP 156, 38042 Grenoble, France}

\author{Anthony J. Horsewill}
\affiliation{School of Physics and Astronomy, University of Nottingham, Nottingham, NG7 2RD}

\author{St{\'e}phane Rols}
\affiliation{Institut Laue-Langevin, BP 156, 38042 Grenoble, France}

\author{Toomas R{\~o}{\~o}m}
\affiliation{National Institute of Chemical Physics and Biophysics, Tallinn, 12618, Estonia}

\author{Richard J. Whitby}
\affiliation{School of Chemistry, University of Southampton,  Southampton, SO17~1BJ, UK}

\author{Malcolm H. Levitt}
\email{mhl@soton.ac.uk}
\affiliation{School of Chemistry, University of Southampton,  Southampton, SO17~1BJ, UK}

\date{\today}

\begin{abstract}
The interactions between atoms and molecules may be described by a potential energy function of the nuclear coordinates. Non-bonded interactions are dominated by repulsive forces at short range and attractive dispersion forces at long range. Experimental data on the detailed interaction potentials for non-bonded interatomic and intermolecular forces is scarce. Here we use terahertz spectroscopy and inelastic neutron scattering to determine the potential energy function for the non-bonded interaction between single He atoms and encapsulating \Csixty fullerene cages, in the helium endofullerenes \HeCsixty{3} and \HeCsixty{4}, synthesised by molecular surgery techniques.  
The experimentally derived potential is compared to estimates from quantum chemistry calculations, and from sums of empirical two-body potentials. 
\end{abstract}

\maketitle

\section{Introduction}

Non-bonded intermolecular interactions determine the structure and properties of most forms of matter. The \emph{potential energy function} specifies the dependence of the potential energy on the nuclear coordinates of the interacting moieties, within the Born-Oppenheimer approximation~\cite{wales_energy_2004}. 
The estimation of potential functions for non-bonded interactions remains an active research area of computational  chemistry~\cite{jensen_introduction_2017,waller_weak_2016,shao_advances_2015}. \textit{Ab initio} methods are capable  of high accuracy but are usually too computational expensive to be applied to anything but very small molecular systems. Computational techniques with good scaling properties such as density functional theory (DFT) are generally imprecise for non-bonded interactions, unless customised adjustments are made~\cite{grimme_consistent_2010,waller_weak_2016,shao_advances_2015}. The accuracy of quantum chemistry algorithms is often assessed by seeking convergence with respect to the calculation level, or number of basis functions~\cite{jensen_introduction_2017}.  

Advances in all fields of science require comparison with experiment. Unfortunately, detailed experimental data on intermolecular potential energy surfaces is scarce. Some information may be gained by comparing crystal structures and energetics with those derived from model potentials~\cite{momany_intermolecular_1974}. The equilibrium structures, dissociation energies and vibrational frequencies of intermolecular complexes and clusters may be studied in the gas phase and molecular beams~\cite{
keutsch_water_2001,
hobza_non-covalent_2009,
zhu_mass_1991,
krause_dissociation_1993,
van_orden_small_1998,
softley_applications_2004}. However these measurements encounter difficulties with control of the local sample temperature, and only provide information on potential minima, and their local properties close to potential minima.
Atomic beam diffraction may also provide information~\cite{farias_atomic_1998,carlos_interaction_1980,cole_probing_1981}.

An ideal set of systems for the study of intermolecular interactions is provided by atomic and molecular endofullerenes, in which single atoms or small molecules are encapsulated in closed carbon cages~\cite{saunders_stable_1993,saunders_noble_1996,komatsu_encapsulation_2005,kurotobi_single_2011}. A range of small-molecule endofullerenes is  available in macroscopic quantities through the multi-step synthetic route known as ``molecular surgery''~\cite{rubin_insertion_2001}, including \HtwoCsixty~\cite{komatsu_encapsulation_2005}, \HtwoCseventy~\cite{murata_synthesis_2008}, \HtwoOCsixty~\cite{kurotobi_single_2011}, \HFCsixty~\cite{krachmalnicoff_dipolar_2016}, \MethaneCsixty~\cite{bloodworth_first_2019}, and their isotopologues. Endofullerenes containing noble gas atoms, and containing two encapsulated species, may also be produced~\cite{murata_encapsulation_2008,murata_synthesis_2008,morinaka_rational_2010,zhang_synthesis_2016,zhang_isolation_2017,bloodworth_synthesis_2020,bacanu_internuclear_2020,hoffman_solid_2021}. Endofullerenes are chemically very stable, may be prepared in a pure and homogeneous solid form, and may be studied at almost any desired temperature. 

\begin{figure*}
	\includegraphics[width=1\linewidth]{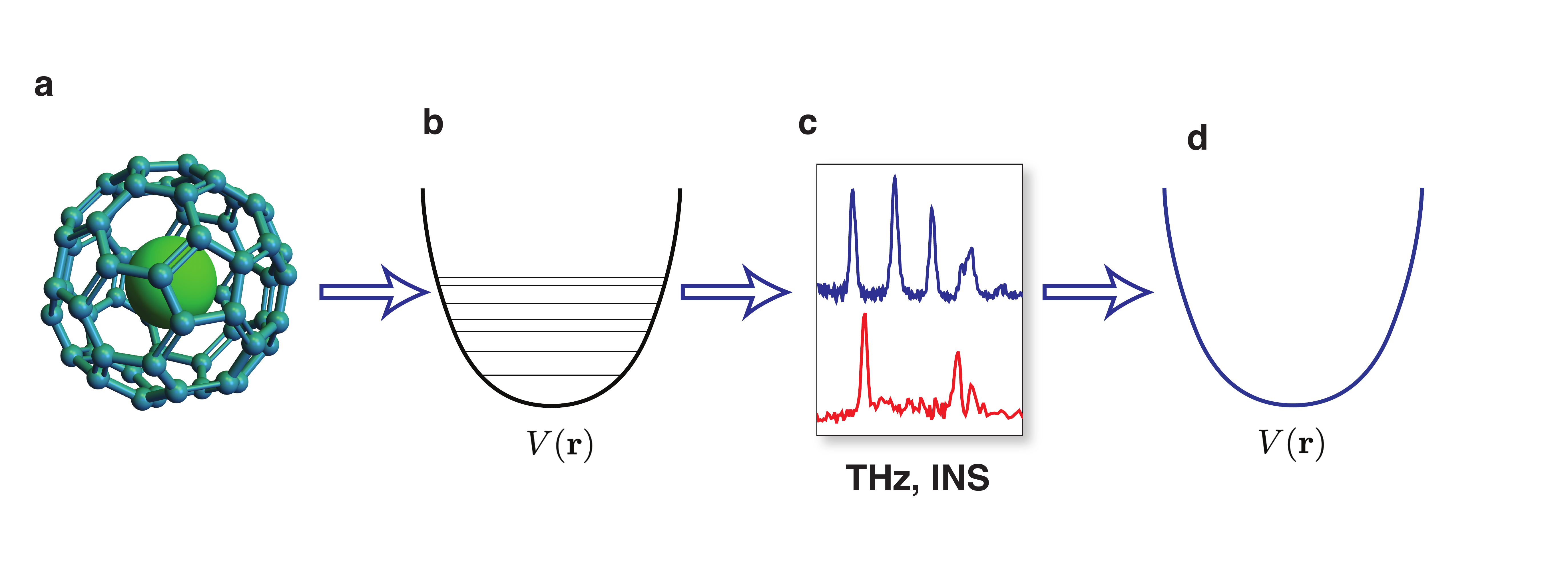}
	\caption{\textbf{(a)} A \Csixty cage encapsulates a single noble gas atom. \textbf{(b)} The confinement potential of the encapsulated atom is described by the function $V(\vecr)$. The quantum energy levels and wavefunctions of the encapsulated atom depend on $V(\vecr)$. \textbf{(c)} Transitions between the energy levels are detected in the bulk solid state at low temperature by terahertz spectroscopy and inelastic neutron scattering. \textbf{(d)} Analysis of the spectroscopic and neutron scattering data allows determination of the potential energy function, which may be compared with computational chemistry estimates. 
		\label{fig:concept}
	}
\end{figure*}

At low temperatures, the translational modes (and for non-monatomic species, the internal degrees of freedom) of the endohedral species are quantized. The quantum levels may be probed by a wide range of spectroscopic techniques ~\cite{levitt_spectroscopy_2013}, including infrared spectroscopy~\cite{mamone_rotor_2009,ge_infrared_2011,beduz_quantum_2012,room_infrared_2013,krachmalnicoff_dipolar_2016,shugai_infrared_2021}, pulsed terahertz spectroscopy~\cite{zhukoV_rotational_2020}, nuclear magnetic resonance~\cite{turro_spin_2009,beduz_quantum_2012,mamone_nuclear_2013,krachmalnicoff_dipolar_2016,bacanu_internuclear_2020,bacanu_fine_2020}, and inelastic neutron scattering~\cite{beduz_quantum_2012,horsewill_inelastic_2012,krachmalnicoff_dipolar_2016,mamone_symmetry-breaking_2016}. When performed at cryogenic temperatures, these techniques reveal a rich energy level structure for the quantized modes of the encapsulated systems~\cite{mamone_rotor_2009,beduz_quantum_2012,horsewill_inelastic_2012,krachmalnicoff_dipolar_2016,mamone_experimental_2016}.%
The quantum structure has been studied in detail using models of the confining potential, sometimes combined with cage-induced modifications of the rotational and vibrational characteristics of the guest molecule~\cite{xu_h-2_2008,xu_quantum_2008,xu_coupled_2009,mamone_rotor_2009,ge_infrared_2011,ge_interaction_2011,mamone_theory_2011,xu_inelastic_2013,room_infrared_2013,mamone_experimental_2016,felker_communication_2016,mamone_experimental_2016,felker_explaining_2017,bacic_perspective_2018,bacic_coupled_2018,xu_endofullerene_2019,felker_flexible_2020,xu_light_2020,shugai_infrared_2021}. 

There are two main ways to describe the interaction potential between the encapsulated species and the cage. One approach describes the interaction potential as a sum over many two-body Lennard-Jones functions involving each endohedral atom and all 60 carbon atoms of the cage~\cite{xu_h-2_2008,xu_quantum_2008,xu_coupled_2009,xu_inelastic_2013,felker_communication_2016,bacic_perspective_2018,bacic_coupled_2018,felker_flexible_2020,xu_light_2020}, sometimes introducing ``additional sites" on the endohedral species as well~\cite{xu_coupled_2009,bacic_perspective_2018,bacic_coupled_2018}. One disadvantage of this approach is that the summed potential has an undesirable dependence on the precise radius of the encapsulating fullerene cage. An alternative approach, which we call ``model-free", describes the interaction potential as a sum of orthogonal spatial functions~\cite{mamone_rotor_2009,ge_infrared_2011,ge_interaction_2011,mamone_theory_2011,room_infrared_2013,mamone_experimental_2016,shugai_infrared_2021}. The latter approach makes no assumptions about the cage geometry and is better-suited for a comparison with computational chemistry methods.  


In this report, we ``go back to basics" by studying the simplest atomic endofullerene, \HeCsixty{}, consisting of \Csixty fullerene cages each encapsulating a single helium atom (figure~\ref{fig:concept}a). Terahertz and neutron scattering data is acquired and fitted by a simple quantum-mechanical model consisting of a particle confined by a three-dimensional potential well. This allows us to define a ``model-free" atom-fullerene potential, with no assumptions about whether it may be expressed as the sum of many two-body interactions. 

Although \HeCsixty{}  was first made in trace amounts by gas-phase methods~\cite{saunders_stable_1993,saunders_noble_1996,giblin_incorporation_1997}, molecular surgery techniques now provide both isotopologues \HeCsixty{3} and \HeCsixty{4} in high purity and macroscopic quantities~\cite{morinaka_rational_2010,hoffman_solid_2021}. These synthetic advances have made it feasible to perform terahertz spectroscopy and inelastic neutron scattering experiments on solid polycrystalline samples of \HeCsixty{} at low temperature, with good signal-to-noise ratio. 

At first sight, \HeCsixty{} is an unpromising object of study by both terahertz spectroscopy and neutron scattering. Since He atoms are neutral, their translational motion is not expected to interact with electromagnetic radiation. Furthermore, both \He{3} and \He{4} isotopes have small neutron scattering cross-sections, and \He{3} is a strong neutron absorber. Fortunately, although these concerns are valid, they are not fatal. The He atoms in \HeCsixty{} acquire a small induced electric dipole through their interactions with the encapsulating cage, and hence interact weakly with the THz irradiation, as in the case of \HtwoCsixty~\cite{mamone_rotor_2009}. The feeble neutron scattering of both He isotopes may be compensated by a sufficiently large sample quantity. 


We compare the experimentally determined potential to estimates from empirical two-body interaction potentials, and from quantum chemistry calculations. Empirical two-body potentials give widely divergent results, even when those potentials are based on experimental helium-graphite scattering data. M{\o}ller-Plesset perturbation theory techniques and density functional theory (DFT) methods which explicitly include, or are empirically corrected to account for, dispersive interactions, are shown to provide good estimates for the interaction potential. 

\section{Materials and Methods}

\subsection{Sample Preparation}
\HeCsixty{3} and \HeCsixty{4} were synthesised using a solid-state process for the critical step, as described in reference~\citenum{hoffman_solid_2021}. The initial filling factors were 30\% to 50\%. The samples were further purified by recirculating HPLC on Cosmosil Buckyprep columns to remove trace impurities of \HtwoOCsixty. Without this precaution, strong neutron scattering by the hydrogen nuclei interferes strongly with the INS measurements. 
For THz spectroscopy samples of high filling factor were required to get sufficient signal and were prepared by further extensive recirculating HPLC. 
All samples were sublimed under vacuum before spectroscopic measurements. 

\subsection{Terahertz Spectroscopy}

THz absorption spectra were measured with an interferometer using a mercury arc light source and a 4~K bolometer as an intensity detector. The typical instrumental resolution was 0.3\wn, which is below the width of the measured THz absorption lines. The \HeCsixty{4} sample had a filling factor of $f=88.2\pm0.5$\% while the \HeCsixty{3} had a filling factor of $f=97.2\pm0.5$\%, as determined by ${}^{13}$C NMR. 
The sample pellets were pressed from fine powders of solid \HeCsixty{}. The temperature dependence of the absorption spectra was measured by using a variable-temperature optical cryostat.
More information is in the Supplementary Material.

\subsection{Inelastic Neutron Scattering}

INS experiments were conducted using the IN1-Lagrange spectrometer at the Institut Laue-Langevin (ILL) in Grenoble. Incident neutrons are provided by the ``hot source`` moderator of the reactor, resulting in a high flux neutron beam. A choice of three different single crystal monochromators, namely Si(111), Si(311) and Cu(220) are used to define the incident energy of the monochromatic neutron beam arriving at the sample using Bragg reflection. The neutrons scattered by the interaction with the sample enter a secondary spectrometer comprising a large area array of pyrolytic graphite analyzer crystals. The focussing geometry of the secondary spectrometer ensures that only neutrons with a fixed kinetic energy of 4.5 meV are detected by the \He{3} detector. INS spectra were recorded in the energy transfer range [5, 200] meV for the \HeCsixty{3} sample, while it was reduced to [5, 60] meV for \HeCsixty{4} as the time allowed for performing the latter experiment was reduced.

The powdered samples, with respective mass of 1067 mg for \HeCsixty{3} ($f=45\%$) 
and 294 mg for \HeCsixty{4} ($f=40\%$) 
were loaded inside an Al foil and further inserted inside a cylindrical annulus before they were mounted at the tip of an orange cryostat and placed inside the IN1 spectrometer beam.
The sample temperature was kept around 2.7~K. In order to subtract background and scattering from Al and from the \Csixty cage, a blank mass matching sample of \Csixty was measured using the same setup and an empty cell was also measured. In order to account for the strong absorption of \HeCsixty{3}, a Cd sample was also measured enabling to correct from the incident energy dependent absorption of the sample. The neutron counts in figure~\ref{fig:INS} were normalized to the incident neutron flux.

\begin{figure}[tbh]
	\centering
	\includegraphics[width=1\linewidth]{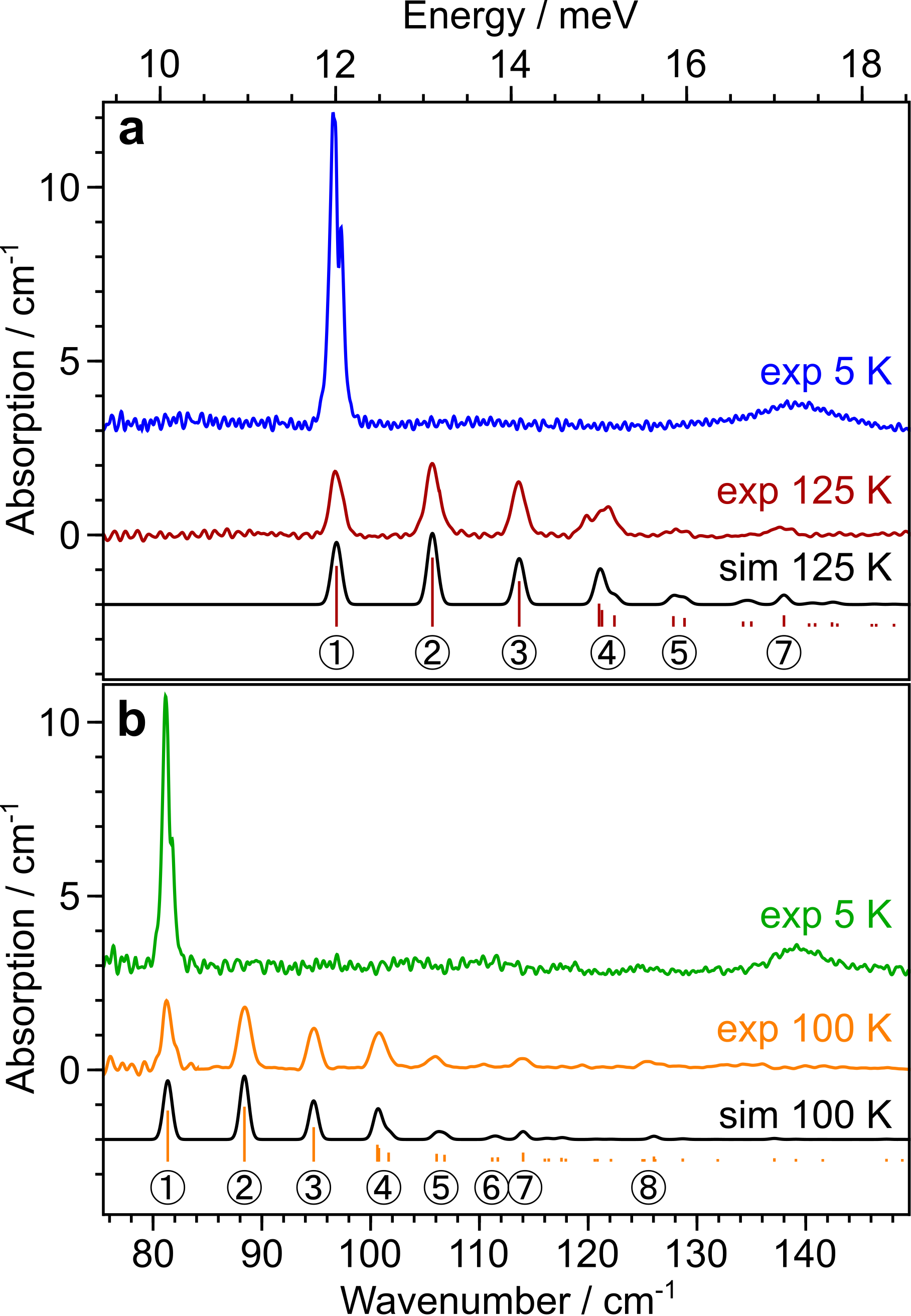}
	\caption{
		THz spectroscopy of He endofullerenes.
		\textbf{(a)} THz absorption spectra of \HeCsixty{3} at temperatures of 5\,K (blue) and 125\,K (red).
		\textbf{(b)} THz absorption spectra of \HeCsixty{4} at temperatures of 5\,K (green) and 100\,K (orange). In both cases, the short vertical bars indicate the predicted positions of the terahertz absorption peaks for the radial potential energy function specified in 
		table~\ref{tab:vvalues}, and their height is proportional to the absorption area. In both cases the black curve is the sum of gaussian peaks with position and area defined by the vertical bars. The THz peaks are numbered according to the transition assignments in figure~\ref{fig:Vbestfit-and-Transitions}(b). 
	}
	\label{fig:THz} 
\end{figure}
\section{Experimental Results}
\subsection{Terahertz spectroscopy}
Terahertz absorption spectra for \HeCsixty{3} and \HeCsixty{4} at two different temperatures are shown in figure~\ref{fig:THz}. For both isotopologues, the high-temperature spectrum displays a comb of several clearly resolved THz peaks, with the \He{3} peaks having higher frequencies than those of \He{4}. As discussed below, the combs of THz peaks indicate that the potential energy function $V(r)$ for the encapsulated He does not have a purely quadratic dependence on the displacement $r$ of the He atom from the cage centre. This indicates that the He dynamics is not well-described as a purely harmonic three-dimensional oscillator. 

The 5\,K spectra in figure~\ref{fig:THz} display a single peak with partially-resolved fine structure, for both \HeCsixty{3} and \HeCsixty{4}. These fundamental peaks correspond to transitions from the quantum ground states of He in the two isotopologues. The fine structure requires further investigation, but may be associated with a small perturbation of the confining potential by the merohedral disorder in the crystal lattice.
Similar effects have been identified for \HtwoCsixty~\cite{mamone_symmetry-breaking_2016}.

\begin{figure}[tbh]
	\centering
	\includegraphics[width=1\linewidth]{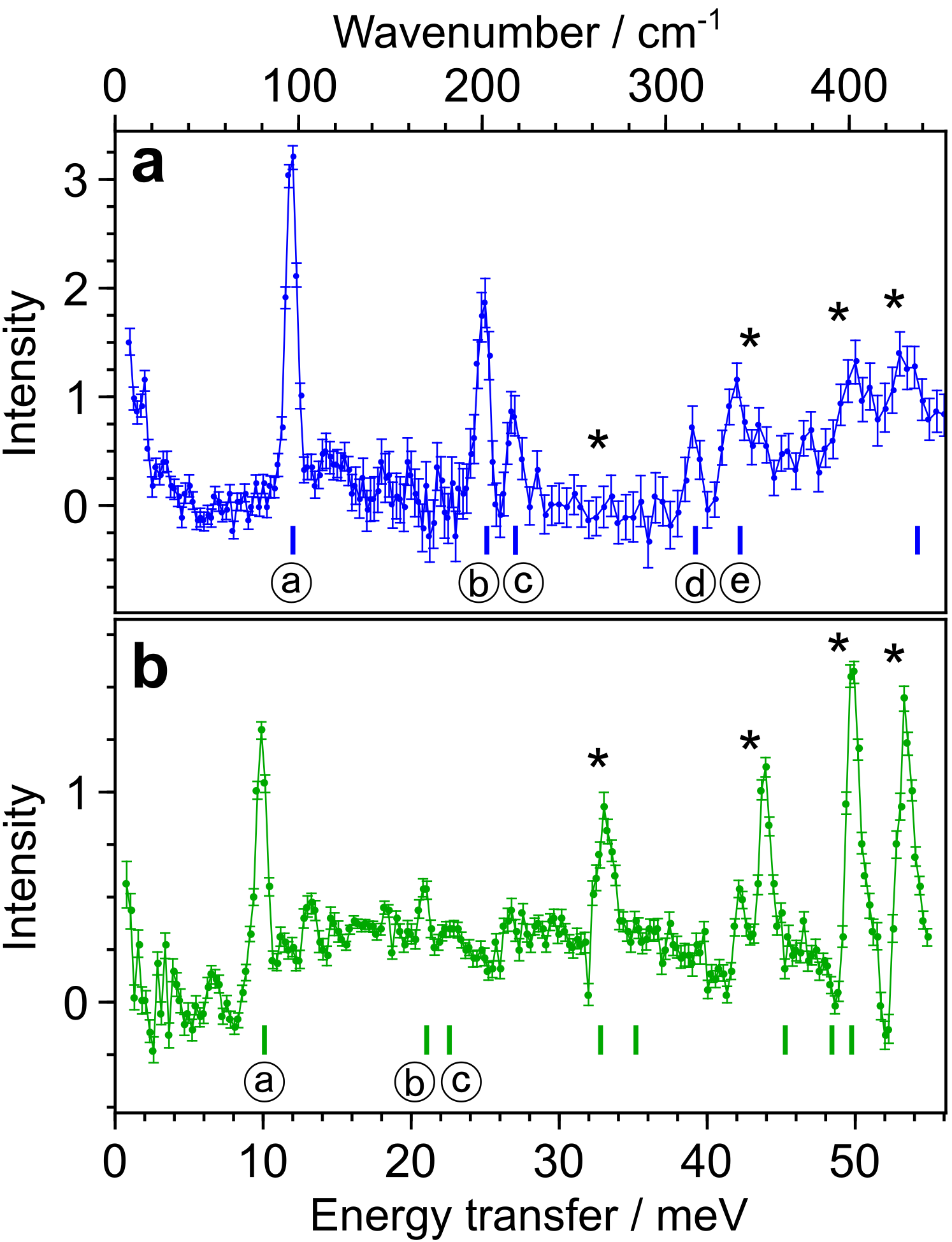}
	\caption{
		Inelastic neutron scattering of He endofullerenes. 
		\textbf{(a)} Inelastic neutron scattering spectra of \HeCsixty{3} at a temperature of 2.7\,K (blue). 
		\textbf{(b)} Inelastic neutron scattering spectra of \HeCsixty{4} at a temperature of 2.7\,K (green). In both cases, a weighted difference between the scattering of \HeCsixty{} and pure \Csixty is shown, with the weighting factors adjusted for best subtraction of the \Csixty background. The short vertical bars indicate the predicted positions of the INS peaks for the quantized He motion under the  radial potential energy function specified in table~\ref{tab:vvalues}. The INS peaks are labelled according to the transition assignments in figure~\ref{fig:Vbestfit-and-Transitions}(b). 
		The peaks above $\sim250\wn$ and marked by asterisks are due to scattering from the \Csixty cages, whose modes are slightly modified in frequency by the presence of endohedral He. 
	}
	\label{fig:INS} 
\end{figure}

\subsection{Inelastic neutron scattering}
Inelastic neutron scattering spectra for \HeCsixty{3} and \HeCsixty{4} are shown in figure~\ref{fig:INS}. The figure shows the difference between the INS of the He endofullerenes and that of pure \Csixty. The INS spectra before subtraction are shown in the Supplementary Material. Since \Csixty has no vibrational modes below $\sim250\wn$, and the low-energy phonon spectrum cancels precisely for the empty and filled fullerenes, the peaks below this energy threshold are clearly attributable to the quantized modes of the confined He atoms.
As in the case of THz spectroscopy, the \He{3} INS peaks are at higher energies than for \He{4}. 

The strong features above $\sim250\wn$ are attributed to the known vibrational modes of \Csixty molecules~\cite{rols_unravelling_2012}. Raman studies have shown that the radial vibrational modes of the \Csixty cages are slightly blue-shifted by the presence of an endohedral noble gas atom~\cite{cimpoesu_vibrational_2011}. These shifts lead to imperfect cancellation in the INS difference spectra, causing the  dispersion-like  features in figure~\ref{fig:INS} which are marked by asterisks. These subtraction artefacts are much stronger for \He{4} than for \He{3}, for two reasons: (i) the \Csixty vibrational modes are slightly more shifted for \He{4} than for \He{3}, due to its larger mass; (ii) \He{4} has a much lower scattering cross-section than \He{3}. 
\begin{figure*}[tbh]
	\centering
	\includegraphics[width=1.0\linewidth]{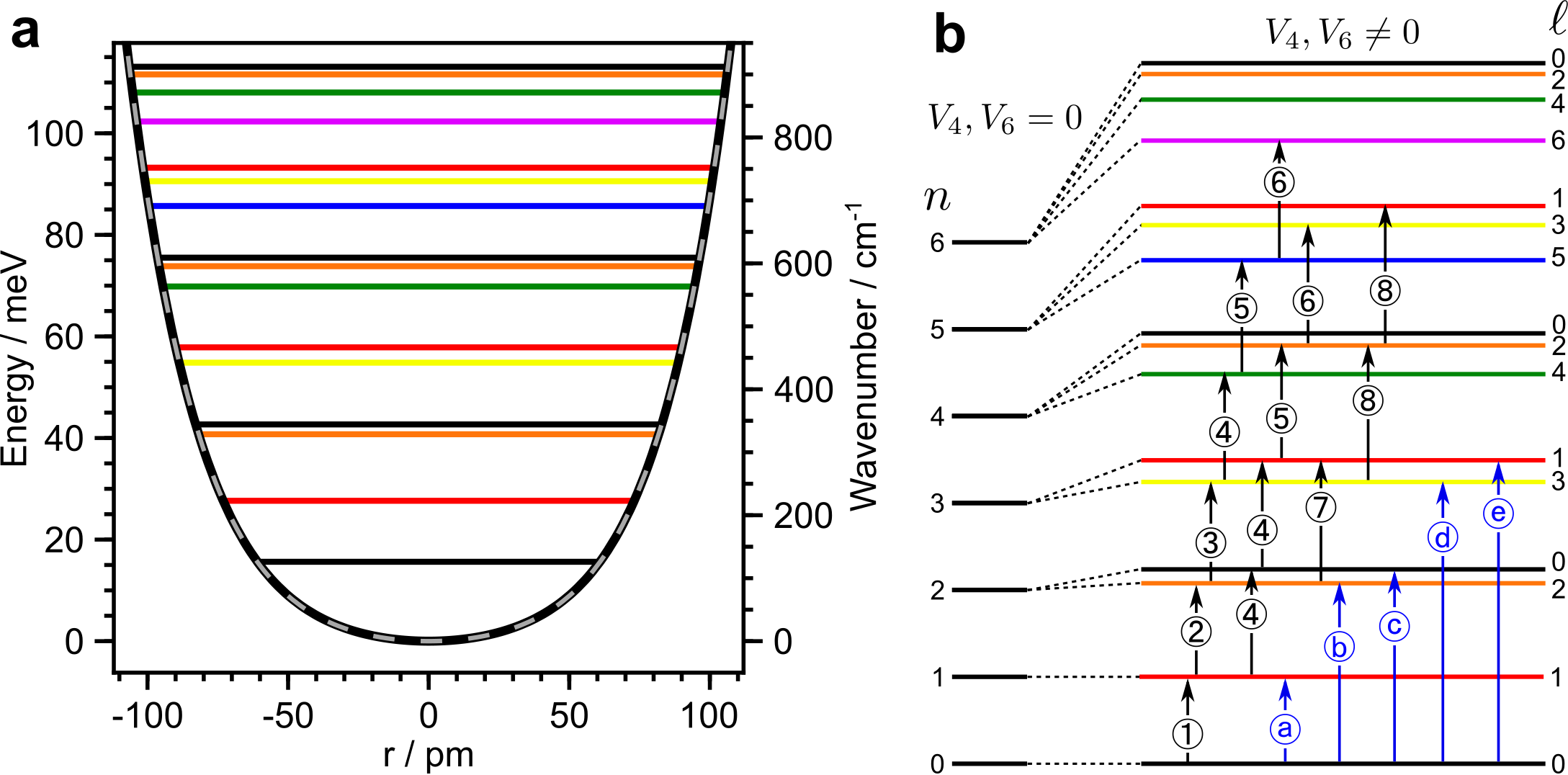}
	\caption{
		\textbf{(a)}
		The radial potential energy functions $V(r)$ for \He{3} in \Csixty (solid black curve) and for \He{4} in \Csixty (dashed grey curve), together with the quantized energy levels for \He{3}. The \He{3} and \He{4} potential curves are superposed within this energy range, leading to a ``railway track" appearance of the plotted curve. The best-fit polynomial  coefficients are given in table~\ref{tab:vvalues}. 
		\textbf{(b)} Energy levels of the confined \He{3} atoms, labelled by the quantum numbers $n\ell$. The energy levels for a harmonic oscillator are shown on the left. The finite $V_4$ and $V_6$ terms break the degeneracies between terms with different $\ell$.  All levels are $(2\ell+1)$-fold degenerate. The transitions observed in THz spectroscopy are labelled by circled numbers in black, and correspond to the peaks in figure~\ref{fig:THz}. The transitions observed in INS are labelled by circled letters in blue, and correspond to the peaks in figure~\ref{fig:INS}. Colours are used to indicate the $\ell$ values of the energy levels.
	}
	\label{fig:Vbestfit-and-Transitions} 
\end{figure*}


\section{Analysis}

\subsection{Energy levels and transitions}
The \Schrodinger equation for the confined atom (within the Born-Oppenheimer approximation), is given by
\begin{equation}\label{eq:SEq}
    \hat{H}(\vecr) \psi_\vecq(\vecr) 
= E_\vecq \psi_\vecq(\vecr)
\end{equation}
where $\vecq$ describes a set of quantum numbers, 
$\vecq=\{q_1,q_2,\ldots\}$, 
and $E_\vecq$ is  the energy of the stationary quantum state. The Hamiltonian operator $\hat{H}$ is given by
\begin{equation}\label{eq:Ham1}
    \hat{H}(\vecr) = - \frac{\hat{p}^2}{2M} + V(\vecr)
\end{equation}
where $\hat{p}$ is the momentum operator and $M$ is the atomic mass. In general, the energy levels $E_\vecq$ and stationary state wavefunctions $\psi_\vecq$ depend strongly on the potential energy function $V(\vecr)$, where $\vecr$ represents the nuclear coordinates of the encapsulated atom (figure~\ref{fig:concept}b).

The potential energy of the He atom inside the cage may be described by a potential function $V(r,\theta,\phi)$, where $r$ is the displacement of the He nucleus from the cage centre, and $(\theta,\phi)$ are polar angles. The \Csixty cage has icosahedral symmetry, but may be treated as spherical to a good approximation, at low excitation energies of the endohedral atom. The angular dependence may be dropped by assuming approximate spherical symmetry, $V(r,\theta,\phi)\simeq V(r)$. We assume a radial potential energy function of the form $V(r) = V_2 r^2 + V_4 r^4 + V_6 r^6$ where $\{V_2,V_4,V_6\}$ are polynomial  coefficients.

The energy eigenvalues and eigenstates are given by $E_{n\ell\ml}$ and $\psi_{n\ell\ml}(r,\theta,\phi)$ respectively. The principal quantum number $n$ takes values $n\in\{0,1,\ldots\}$ with the angular momentum quantum number $\ell$ given by $\ell\in\{0,2,\ldots n\}$ (for even $n$) and $\ell\in\{1,3,\ldots n\}$ (for odd $n$)~\cite{cohen-tannoudji_quantum_2020}. 
The azimuthal quantum number takes values $\ml\in\{-\ell,-\ell+1,\ldots +\ell\}$.
%
For spherical symmetry, the energies are independent of $\ml$, so the energy level $\Enl$ is $(2\ell+1)$-fold degenerate. The stationary quantum states  $\PSInlm (r,\theta,\phi)$ are given by products of radial functions $\Rnl (r)$ and spherical harmonics $\Ylm (\theta,\phi)$, just as for the electronic orbitals of a hydrogen atom~\cite{cohen-tannoudji_quantum_2020}.

The eigenvalues and eigenstates depend on the potential coefficients $\{V_2,V_4,V_6\}$ and the mass of the He atom. The electric-dipole-allowed transitions, which are observed in THz spectroscopy and described by the induced dipole moment coefficient $A_1$, have the selection rule $\Delta\ell=\pm1$, see Supplementary Material. 
There are no relevant selection rules for the neutron scattering peaks. 

\subsection{Fitting of the Potential}
We treat the $V_4$ and $V_6$ terms as perturbations of the quadratic $V_2$ term, which corresponds to an isotropic three-dimensional harmonic oscillator. The solutions of the \Schrodinger equation for the isotropic 3D harmonic oscillator are well-known~\cite{shaffer_degenerate_1944,cohen-tannoudji_quantum_2020}, and are given by:
\begin{equation}\label{eq:eigenstate_3D}
\ket{n\ell m}(r,\theta,\phi)  = R_{n\ell}(r)\SpherHar{}{\ell m}(\theta,\phi), 
\end{equation}
where the principal quantum number is given by $n\in\{0,1,2,\ldots\}$ and the angular momentum quantum number $\ell$ takes values $\{0,2,\ldots n\}$ for even $n$, and $\{1,3,\ldots n\}$ for odd $n$. The radial wavefunctions $R_{n\ell}(r)$ are proportional to generalised Laguerre polynomials~
\cite{flugge_practical_1999,bransden_quantum_2000}, while the angular parts $\SpherHar{}{\ell m}$ are spherical harmonics. The energy eigenvalues are given by
\begin{equation}
  E_{n\ell\ml}=\hbar \omega_0 (n + \frac{3}{2})
\end{equation}
with the fundamental vibrational frequency $\omega_0=(2V_2/\mu)^{1/2}$, where $\mu$ is the reduced mass (assumed here to be equal to the mass of the \He{3} or \He{4} atom, since each \Csixty molecule is more than two orders of magnitude more massive than the encapsulated atom, and is also coupled to the lattice). 

The \Schrodinger equation was solved approximately for finite $V_4$ and $V_6$ by numerically diagonalizing a matrix with elements given by
$\bra{n\ell m}V_4 r^4 + V_6 r^6 \ket{n'\ell'm'}$. 
Since the assumed Hamiltonian retains isotropic symmetry, all matrix elements are independent of the quantum number $m$ and vanish for $\ell\neq\ell'$ and $m\neq m'$. 
In practice the matrix was bounded by quantum numbers  $n\leq18$, after checking for convergence. 
The THz peak intensities and peak positions were fitted, as described in the Supplementary Material,  allowing numerical estimation of the potential parameters $V_2$ (or $\omega_0$), $V_4$ and $V_6$, and the induced dipole moment amplitude $A_1$. The derived eigenvalues  were used to estimate the INS peak positions. 
\begin{table}[bt]
	\centering
	\caption{
		Best fit polynomial coefficients and confidence limits for the radial potential  function $V(r)=V_2 r^2 + V_4 r^4 + V_6 r^6$ and induced dipole  
		function $d_{1q}  =
		\sqrt{4\pi/3} \;
		A_{1} \, r \, \SpherHar{}{1q}(\theta,\phi)$ experienced by the confined He isotopes, see SI.
	}
	\begin{ruledtabular}
		\begin{tabular}{lcc}
			
			Parameter & \Helium{3} & \Helium{4} \\
			\hline
			$V_2$ / \meVperpmsq & $(2.580 \pm 0.011)\; 10^{-3}$ & $(2.4998 \pm 0.0016)\; 10^{-3}$ \\  
			$V_4$ / \meVperpmfo &  $(3.370 \pm 0.060)\; 10^{-7}$ &   $(3.610 \pm 0.060)\; 10^{-7}$\\ 
			$V_6$ / \meVperpmsx &  $(2.786  \pm 0.005)\; 10^{-11}$ &  $(2.634 \pm 0.021)\; 10^{-11}$\\
			$A_{1}$ / D pm$^{-1}$ & $(4.38 \pm 0.04)\; 10^{-4}$ & $(4.58 \pm 0.06)\; 10^{-4}$\\ 
		\end{tabular}
	\end{ruledtabular}
	\label{tab:vvalues}
\end{table}

The fitting of the potential was performed independently for the two He isotopes. The best fit solutions for the potential coefficients, and their confidence limits, are given in table~\ref{tab:vvalues}.

Figure~\ref{fig:Vbestfit-and-Transitions}(a) shows the best-fit potential functions for \He{3} and \He{4} inside the interior cavity of \Csixty. The best-fit potential has a distinct U-shape  which deviates strongly from the parabolic form of a harmonic oscillator. The best-fit potential curves \He{3} and \He{4} are indistinguishable within the plotted energy range.

An energy level diagram for the confined He atoms, marked with the observed transitions, is shown in figure~\ref{fig:Vbestfit-and-Transitions}(b). The predicted positions of the relevant THz and INS transitions are shown by the vertical bars in figures~\ref{fig:THz} and \ref{fig:INS}. Although some of the higher-energy transitions in the INS data are partially obscured by \Csixty features, the agreement with the spectroscopic results is gratifying. The close correspondence of the derived potential curves for \He{3} and \He{4}, despite the different masses of the isotopes and the very different observed frequencies, attests to the validity of the determination of $V(r)$. 

\subsection{Comparison with Empirical Potentials}
\begin{figure}[tb]
	\centering
	\includegraphics[width=1.0\linewidth]{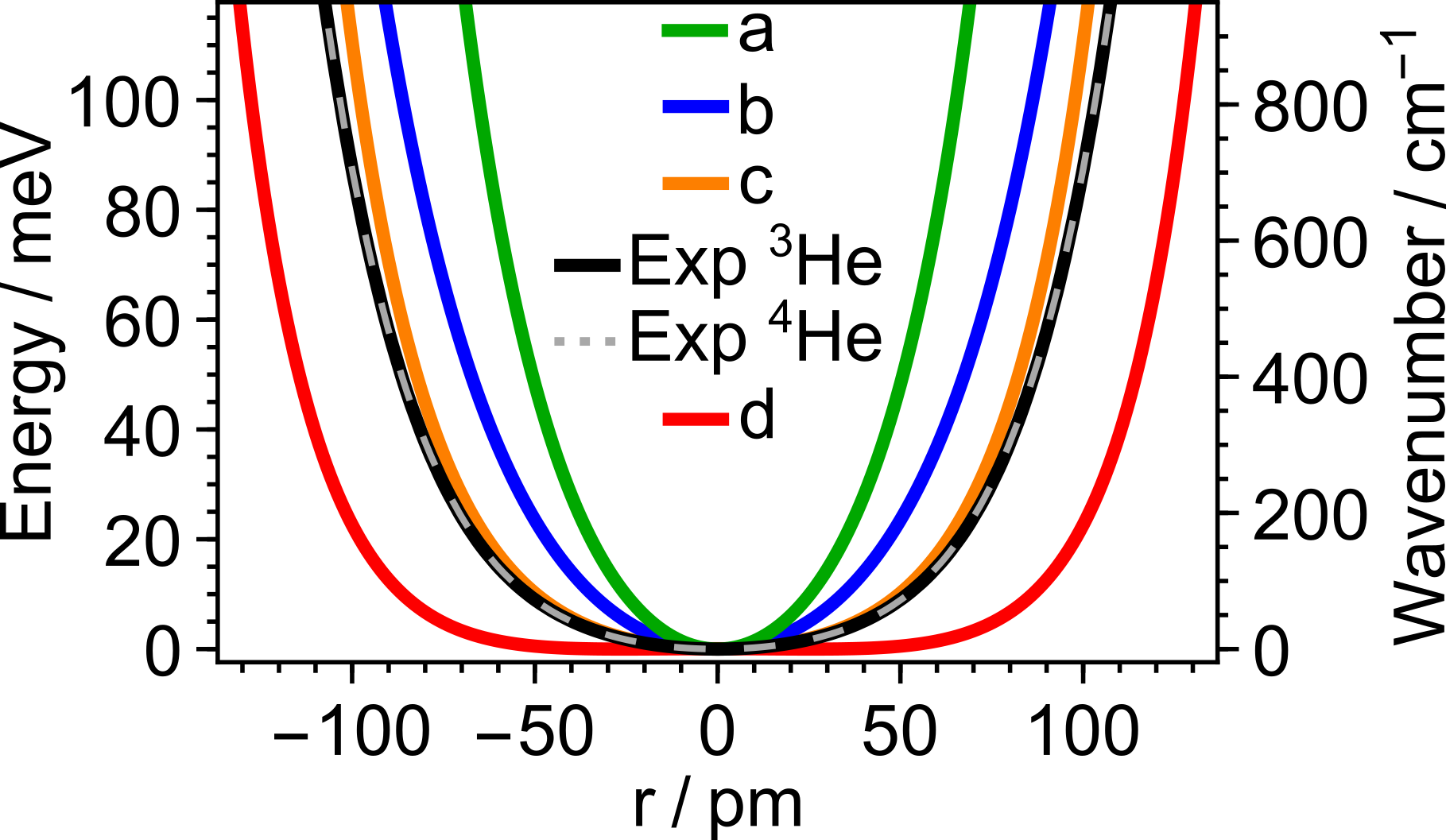}
	\caption{Comparison of the experimentally determined radial potentials $V(r)$ (\He{3}: solid black curve; \He{4}: dashed grey curve, superposed on the \He{3} curve to give a ``train track" appearance) 
		with sums of reported He$\,\cdots\,$C interaction potentials: 
		\textbf{(a, green)} Lennard-Jones 6-8-12 potential with parameters from Carlos \textit{et al.}~\cite{carlos_interaction_1980};
		\textbf{(b, blue)} Modified Buckingham potential (implemented in the MM3 program, as reported by Jim{\'e}nez-V{\'a}zquez \textit{et al.}~\cite{jimenezvazquez_equilibrium_1996}); \textbf{(c, orange)} Lennard-Jones 6-12 potential with parameters from Pang and Brisse~\cite{pang_endohedral_1993}; \textbf{(d, red)} Lennard-Jones 6-12 potential with parameters from Carlos \textit{et al.}~\cite{carlos_interaction_1980}; The potentials used in (a) and (d) were used for the fitting of He$\,\cdots\,$C scattering data~\cite{carlos_interaction_1980}. The functional forms of the potentials and their associated parameters are given in the Supplementary Material. In all cases the He atom was displaced from the cage centre towards the nucleus of a carbon atom. The confidence limits in the structural data for \Csixty~\cite{leclercq_precise_1993} lead to error margins on the empirical curves which are smaller than the plotted linewidths. 
	}
	\label{fig:EmpiricalPotentialsComparison} 
\end{figure}
There have been numerous attempts to model the non-bonded interactions between atoms using empirical two-body potential functions such as the Lennard-Jones (LJ) 6-12 potential, or by more complex functional forms. Suitable functions and parameters have been proposed for the He$\,\cdots\,$C interaction~\cite{terry_amos_atomatom_1990,pang_endohedral_1993,jimenezvazquez_equilibrium_1996, carlos_interaction_1980,cole_probing_1981}. Some of the proposed two-body potentials were developed for modelling the scattering of He atoms from a graphite surface~\cite{carlos_interaction_1980,cole_probing_1981}. 

Figure~\ref{fig:EmpiricalPotentialsComparison} compares the experimental $V(r)$ curve with predictions from published He$\,\cdots\,$C two-body interaction functions.  In each case, the total potential energy $V(r)$ was estimated by locating the He atom a distance $r$ along a line from the centre of the cage towards a C atom, and summing the contributions from all 60 two-body He$\,\cdots\,$C potentials. The direction of the He displacement has a negligible effect on the calculated potential curves over the relevant energy range (see Supplementary Material). The derived potentials are very sensitive to the geometry of the \Csixty cage, especially its radius $R$. We fixed the locations of all C nuclei to the best current estimates from neutron diffraction~\cite{leclercq_precise_1993}, as follows: Bond lengths $h= 138.14 \pm 0.27\picom$ for C-C bonds shared by two hexagons, $p= 145.97 \pm 0.18\picom$ for C-C bonds shared by a hexagon and a pentagon, and distance of all carbon atoms from the cage centre $ R = 354.7 \pm 0.5\picom$.  The width of the curves in figure~\ref{fig:EmpiricalPotentialsComparison} is greater than their confidence limits, which are dominated by the uncertainties in the structural parameters. Explicit functional forms and parameters for the empirical two-body potentials are given in the Supplementary Material. 

The most striking feature of Figure~\ref{fig:EmpiricalPotentialsComparison} is the wide variation of derived potentials for different two-body interaction models. Of all the proposed two-body potentials, the Lennard-Jones 6-12 potential with parameters given by Pang and Brisse~\cite{pang_endohedral_1993} (curve a) provides the best agreement with experiment. The isotropic two-body potentials derived by fitting experimental He/graphite scattering data~\cite{carlos_interaction_1980,cole_probing_1981} (curves c and d) give poor fits to the experimental \HeCsixty{} potential. 
\subsection{Comparison with Quantum Chemistry }
The \HeCsixty{} system is too large to be treated at the full \emph{ab initio} level of quantum chemistry. The availability of an experimental radial potential function $V(r)$ allows the direct evaluation of approximate computational chemistry techniques -- not only at the equilibrium geometry, but also for displacements of the He atom from the centre of the \Csixty cage. 


The radial potential $V(r)$ was evaluated by estimating the energy of a \HeCsixty{} system using a range of computational chemistry algorithms, with the He atom displaced by $r$ from the centre of the \Csixty cage. In all cases the locations of the carbon atoms were fixed to the \Csixty geometry as determined by neutron diffraction~\cite{leclercq_precise_1993}, with the same parameters as used for the evaluation of the empirical potentials. The He was moved on the line connecting the cage centre to a carbon nucleus. The direction of the He displacement has a negligible effect on the predicted potential curves over the relevant energy range (see Supplementary Material). 
The potentials were calculated using the Psi4 program~\cite{psi4}. The functionals used for DFT were:
(i) the B3LYP functional, which is one of the most popular semi-empirical hybrid functionals~\cite{mardirossian2017thirty, lee_development_1988,vosko_accurate_1980,becke_densityfunctional_1993,stephens_ab_1994}; (ii) the B3LYP functional including the Grimme D3 empirical dispersion correction with Beck-Johnson damping~\cite{grimme_consistent_2010,grimme_effect_2011}; (iii) the $\omega$B97X-V functional, which includes a contribution from the non-local VV10 correlation functional and is designed to handle non-covalent interactions~\cite{mardirossian2017thirty}. 
The potential was also calculated using second-order M{\o}ller-Plesset perturbation (MP2) theory~\cite{jensen_introduction_2017} including empirical spin-component-scaling factors (SCS)~\cite{SCS}. 
All potential calculations employed a counterpoise basis-set-superposition-error correction, and converged to a good approximation with the correlation-consistent cc-pVXZ (X=D, T, Q, 5) basis sets~\cite{dunning1989gaussian,woon1994gaussian}. 
More details on the quantum chemistry calculations are given in the Supplementary Material.

\begin{figure}[bt]
	\centering
	\includegraphics[width=1.0\linewidth]{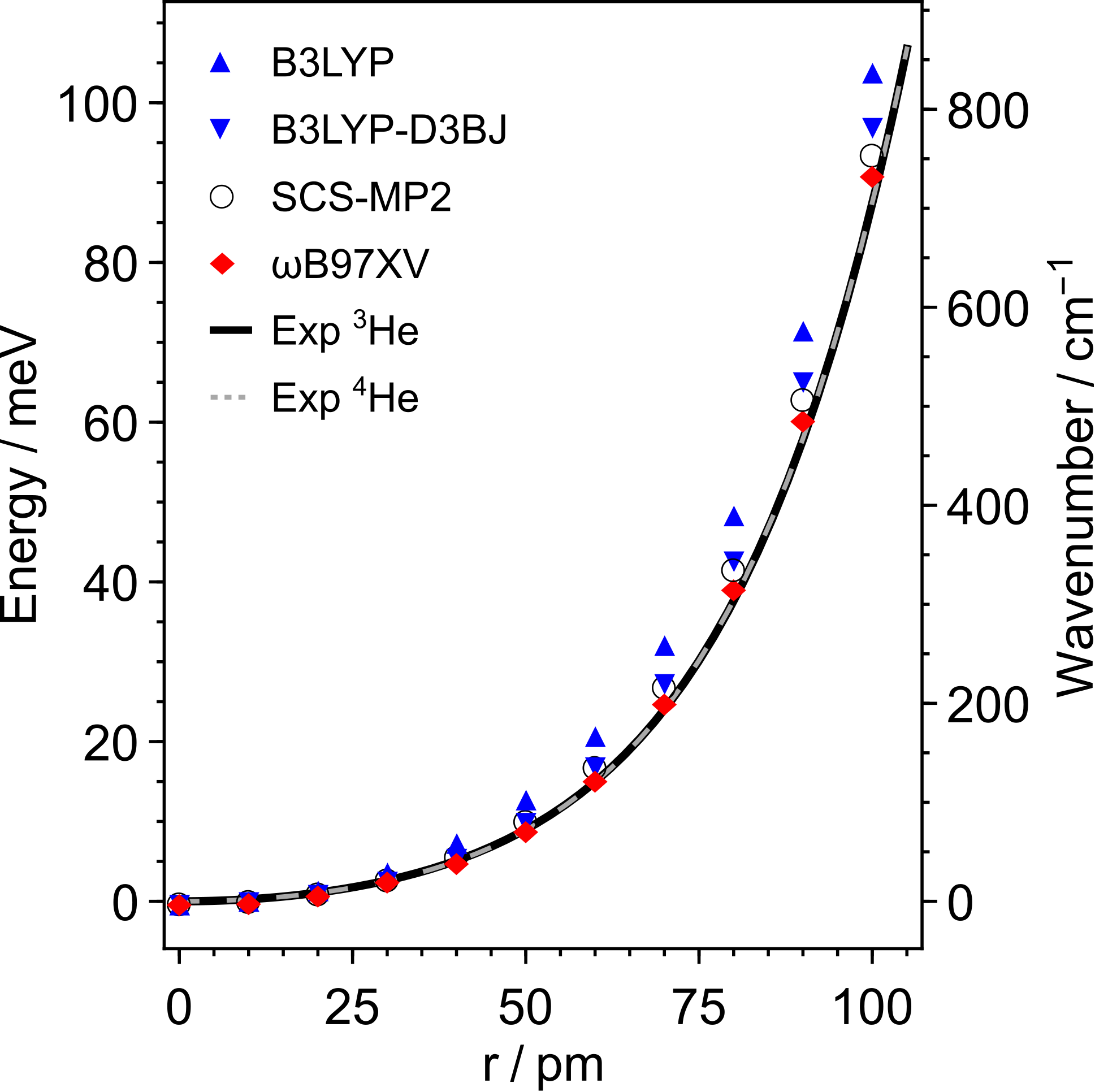}
	\caption{
		Comparison of the experimentally determined \HeCsixty{} radial potentials $V(r)$ (\He{3}: solid black curve; \He{4}: dashed grey curve, superposed on the \He{3} curve)
		with quantum chemical calculations using density functional and M{\o}ller-Plesset perturbation theories~\cite{jensen_introduction_2017}, as follows: (\blue{$\blacktriangle$}) DFT using the B3LYP functional~\cite{lee_development_1988,vosko_accurate_1980,becke_densityfunctional_1993,stephens_ab_1994}; (\blue{$\blacktriangledown$}) DFT using the B3LYP functional with D3BJ correction~\cite{grimme_consistent_2010,grimme_effect_2011}; (\textcolor{red}{\rotatebox[origin=c]{90}{$\blacklozenge$}}) DFT using the $\omega$B97XV functional~\cite{wB97XV,mardirossian2017thirty}; (\scalebox{1.5}{$\circ$}) Spin-component-scaled M{\o}ller-Plesset perturbation theory (SCS-MP2).~\cite{SCS}
	}
	\label{fig:QChem} 
\end{figure}

Some comparisons are shown in Figure~\ref{fig:QChem}. Density functional theory with the popular B3LYP functional~\cite{mardirossian2017thirty,lee_development_1988,vosko_accurate_1980,becke_densityfunctional_1993,stephens_ab_1994} overestimates the steepness of the confining potential, although the correspondence with experiment is improved by including the empirical D3 correction with Beck-Johnson damping~\cite{grimme_consistent_2010,grimme_effect_2011}. DFT with the $\omega$B97X-V functional~\cite{mardirossian2017thirty}, and M{\o}ller-Plesset perturbation (MP2) theory with spin-component-scaling factors (SCS)~\cite{SCS}, both give an acceptable correspondence between the calculated and experimentally determined potentials. 


\section{Discussion}

We have showed that the quantized energy levels of helium atoms encapsulated in \Csixty cages may be probed by THz spectroscopy and INS, despite the weak interactions of the He atoms with the electromagnetic field and with neutrons. The spectroscopic features were analysed to obtain a detailed potential energy function for the interaction between the encapsulated species and the surrounding cage -- an interaction dominated by non-bonded dispersion forces which are hard to estimate experimentally. An excellent correspondence was obtained between the interaction potentials derived from independent \HeCsixty{3} and \HeCsixty{4} measurements, despite the different peak positions for the two samples. 

The experimental $V(r)$ curve was compared with sums of published two-body He$\,\cdots\,$C interactions. With few exceptions the summed two-body potentials have a poor correspondence with the experimental result. It is not a great surprise that the interaction of a He atom with a highly delocalized electronic structure such as \Csixty is hard to model as the sum of individual atom-atom interactions.

We also compared the experimentally derived interaction potential with those derived by quantum chemistry techniques. This allowed the validation of DFT methods which have been developed to deal with dispersive interactions, including the popular B3LYP functional with the D3 empirical dispersion correction~\cite{grimme_consistent_2010,grimme_effect_2011}, and the $\omega$B97X-V functional which incorporates the non-local VV10 correlation functional and has been parameterised using a training set rich in non-bonding interactions~\cite{mardirossian2017thirty}. M{\o}ller-Plesset perturbation theory with spin-component-scaling factors~\cite{SCS} also provides a good description of the confining potential of the encapsulated He atoms.

There are small discrepancies between the calculated and observed potentials. However it is not yet known whether the remaining discrepancies reflect the limitations in the quantum chemistry algorithms, or the limitations in the assumptions made when interpreting the experimental data -- for example, the neglect of the influence exerted by the encapsulated He atoms on the cage radius. Precise measurements of the \HeCsixty{} cage geometry by neutron scattering or X-ray diffraction are planned.

He atoms are small, have no static dipole moment, and a low polarizability. This makes \HeCsixty{} a relatively easy case for computational chemistry. A stiffer challenge for computational chemistry is likely to be presented by compounds in which the endohedral species is polar, such as \HtwoOCsixty~\cite{murata_encapsulation_2008} and \HFCsixty~\cite{krachmalnicoff_dipolar_2016}, and by endofullerenes such as \MethaneCsixty~\cite{bloodworth_first_2019}, where the fit with the cage is much tighter. Furthermore, the study of systems with multiple atoms or molecules encapsulated in the same fullerene cage~\cite{murata_encapsulation_2008,zhang_synthesis_2016,zhang_isolation_2017} should allow the study of non-bonded molecule-molecule and molecule-atom interactions. 

\section*{Author contributions}
G.R.B. conceived the THz experiments.
M.W., G.H. and R.J.W. synthesised and purified the compounds.
G.R.B., T.J., A.S., U.N. and T.R. performed the THz experiments and processed the THz data. 
A.J.H. and S.R. designed the INS experiments. 
G.R.B., M.A. and S.R. performed the INS experiments and processed the INS data. 
G.R.B., M.A. and T.R. derived the potential function.
J.R. and R.J.W. performed the quantum chemistry calculations.
G.R.B., M.A., J.R., A.J.H., S.R., T.R., R.J.W. and M.H.L. developed the concept and drafted the paper. All authors reviewed the manuscript.

\begin{acknowledgments}
This research was supported by EPSRC-UK (grant numbers EP/P009980/1, EP/T004320/1 and EP/P030491/1), the Estonian Ministry of Education and Research institutional research funding IUT23-3, personal research funding PRG736,  the European Regional Development Fund project TK134, and the European Union’s Horizon 2020 research and innovation programme under the Marie Sk{\l}odowska-Curie grant agreement No 891400. The Institut Laue-Langevin is acknowledged for providing neutron beam time, and support for M.A. through the ILL PhD program.
The authors acknowledge the use of the IRIDIS High Performance Computing Facility, and associated support services at the University of Southampton, in the completion of this work.
\end{acknowledgments}

\section*{Data Availability Statement}

The data that support the findings of this study are available from the corresponding author upon reasonable request.

%

\end{document}


\maketitle


\newpage



\section*{THz spectroscopy}%
    The THz transmission spectra were obtained using the Bruker interferometer Vertex 80v and an optical cold-finger type continuous flow cryostat with two thin-film polypropylene windows.
    For the measurements the powder of resublimed \HeCsixty{} samples was pressed under vacuum into a 3\,mm diameter hole of a brass frame. 
    The mass and thickness of \HeCsixty{3} sample were  28\,mg and 2.16\,mm and of \HeCsixty{4} sample 21\,mg and 1.72\,mm.
    The brass frame with the pellet was inserted into a sample chamber with two thin-film polypropylene windows and with  a vacuum  line for  pumping and filling with Helium heat exchange gas. 
    The sample chamber was in a thermal contact with  the cold finger of the cryostat.
    The cryostat was placed inside the  Vertex 80v sample compartment. 
    The cold finger was moved up and down by letting the beam through the sample chamber or through a reference hole with 3\,mm diameter. 
    The mercury arc lamp, 6\,$\mu$m Mylar beamsplitter and 4\,K bolometer were used to record the transmission spectra below 300\wn.
    The apodized resolution was typically 0.3\wn or better.
    
    The transmission $ T_r(\omega)$ was measured as the light intensity transmitted by the sample chamber divided by the light intensity transmitted by the reference hole.
    The absorption coefficient $ \alpha(\omega) $ was calculated from the transmission $ T_r(\omega)$ through $\alpha(\omega)= -d^{-1} \ln\left[ T_r(\omega)R_{\mathrm{corr}}^{-1}\right]$, with $R_{\mathrm{corr}}$ as the amount of radiation reflected from the sample pellet surface and from the windows of the sample chamber. $R_{\mathrm{corr}}$ adds to the background absorption but does not affect the intensities of  \HeCsixty{} absorption lines.
    
    Spectra presented in the paper were measured at low temperature, 5~K, and at high temperature, 100~K for \HeCsixty{4} and 125~K for \HeCsixty{3}.
    The low and high $T$ absorption spectra were treated differently before fitting the peaks with Gaussian functions.
    The baseline correction by subtracting  the slowly  changing background was performed on the low  temperature absorbance spectra  and on the lowest frequency line in the high $T$ spectra.
    The baseline of the high  $T$ spectrum above the lowest frequency resonance was corrected by subtracting the 5~K spectrum from the high $T$ spectrum.
    The position of a broad line observed at 140\wn in the 5~K spectra is independent of the  mass of He and independent of $T$ as it  subtracts out  in the high $T$ spectra if using  the 5~K spectrum  as a baseline.
    Therefore, the 140\wn resonance is not caused  by the presence of endohedral He and is likely a \Csixty lattice mode.
    
    \subsection*{\HeCsixty{} filling factors from \Cth NMR}
    The \Cth solution NMR spectra of \HeCsixty{3} and \HeCsixty{4}, in \ODCBd4 (Sigma-Aldrich), were acquired in order to measure the filling factors of the endofullerenes. The measurements were performed at 298\kel and a field of 16.45\T on a Bruker Ascend 700 NB magnet fitted with a Bruker TCI prodigy 5 mm liquids cryoprobe and a Bruker AVANCE NEO console.
    
\begin{figure}[hbt!]
          \centering
          \includegraphics[width=1.0\textwidth]{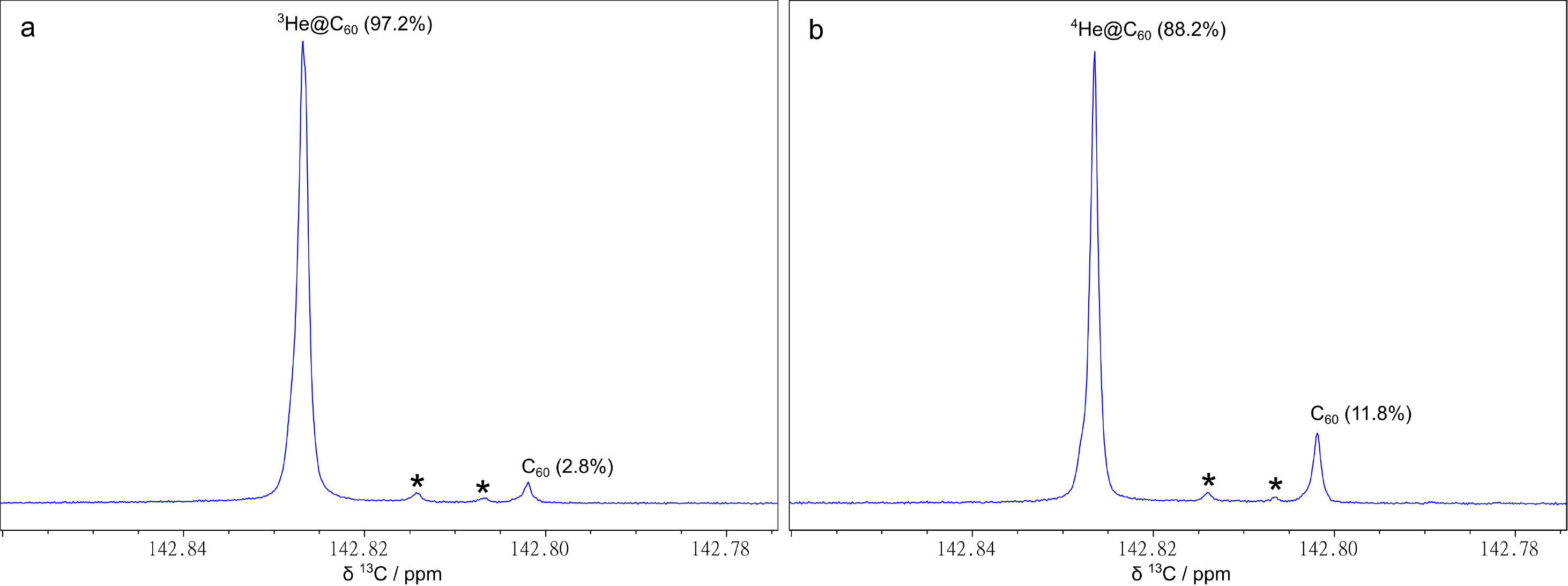}
          \caption{\Cth solution NMR spectra of \HeCsixty{} in \ODCBd4 at 298\kel and 16.45\T. \textbf{(a)} \HeCsixty{3} $\sim$9.9 mM (f $ =97.2\pm0.5\%$) acquired with 720 transients. \textbf{(b)} \HeCsixty{4} $\sim$6.3 mM (f $ =88.2\pm0.5\%$)  acquired with 424 transients.
          The side peaks of \HeCsixty{3} and \HeCsixty{4} are marked with an asterisks; they arise from \Cthtwo isotopomers of \Csixty, see reference \citenum{bacanu_fine_2020} for more details.
          }
          \label{fig:3HeC60_4HeC60_13Cspc_THz_ff_TEST_298K_700MHz}
\end{figure}

    The \Cth solution NMR spectrum of \HeCsixty{3} is shown in fig.~\ref{fig:3HeC60_4HeC60_13Cspc_THz_ff_TEST_298K_700MHz} (a), which results in a filling factor f $ =97.2\pm0.5$\%.
    The \Cth solution NMR spectrum of \HeCsixty{4} is shown in fig.~\ref{fig:3HeC60_4HeC60_13Cspc_THz_ff_TEST_298K_700MHz} (b), which results in a filling factor f $ =88.2\pm0.5$\%.

\section*{Inelastic neutron scattering}

    \subsection*{\HeCsixty{3} INS}
\begin{figure}
          \centering
          \includegraphics[width=0.8\textwidth]{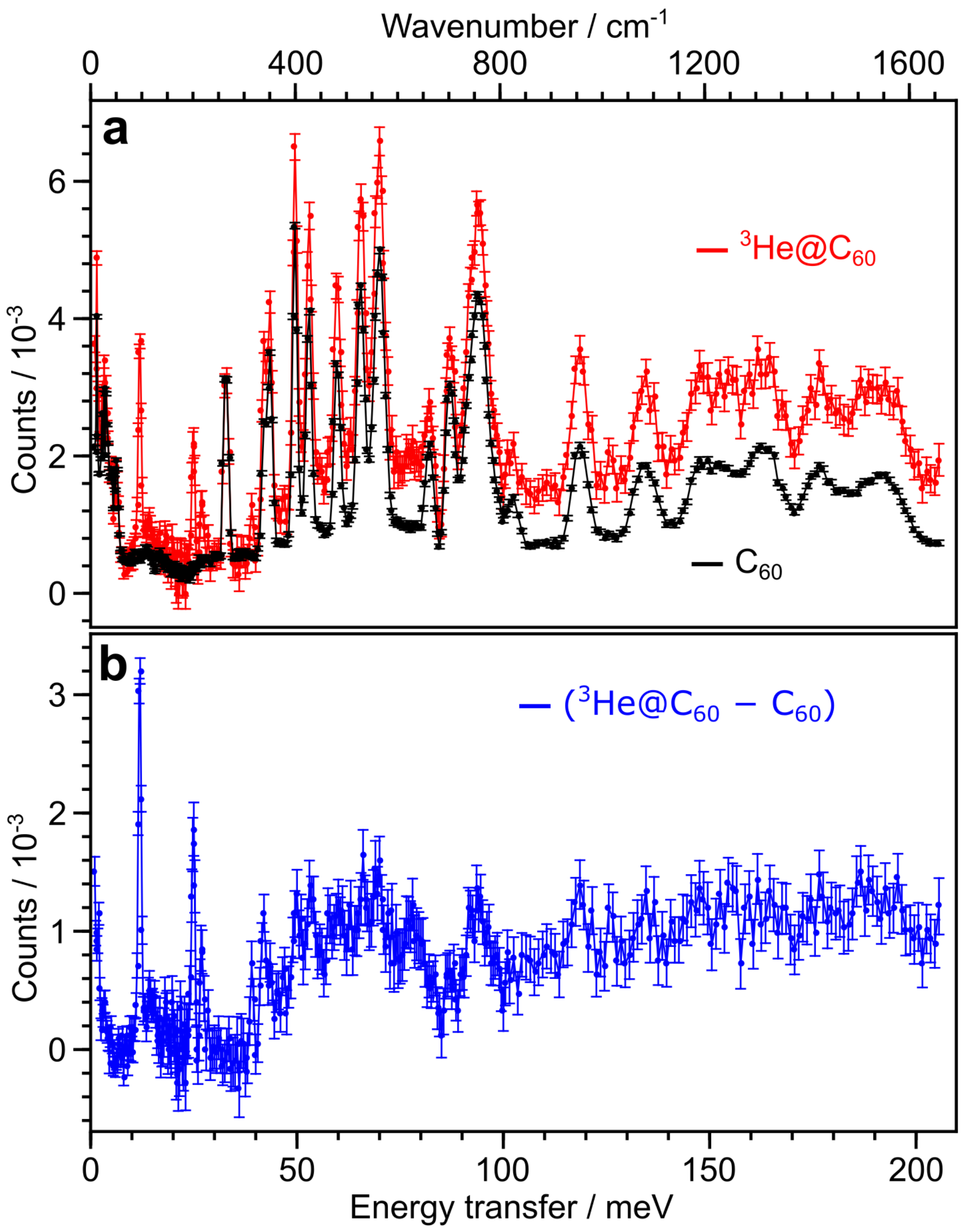}
          \caption{\textbf{(a)} IN1 LAGRANGE INS (transmission corrected) spectra of: 1067 mg \HeCsixty{3} (f=45\%) in red and 1067 mg \Csixty in black, at 2.7\kel. \textbf{(b)} in blue the difference \HeCsixty{3}$-$\Csixty, red$-$black, from (a) is shown.
          Counts are normalised with respect to the monitor neutron count.
          %
          }
          \label{fig:3HeC60_INS_IN1_fig_tCORR_diffBLU_3HeC60RED_C60x0.154BLK_FIG}
\end{figure}

     The low temperature INS spectra of 1067 mg \HeCsixty{3} (f=45\%) and 1067 mg \Csixty are shown in fig.~\ref{fig:3HeC60_INS_IN1_fig_tCORR_diffBLU_3HeC60RED_C60x0.154BLK_FIG} (a), in red and black respectively.
     
 The \HeCsixty{3} spectrum is corrected for the neutron absorption by the encapsulated \He{3}. The \Csixty spectrum is scaled to best match the \HeCsixty{3} in the low energy transfer region of the spectrum, in order to obtain optimum background subtraction. In fig.~\ref{fig:3HeC60_INS_IN1_fig_tCORR_diffBLU_3HeC60RED_C60x0.154BLK_FIG} (b) the difference between \HeCsixty{3} and \Csixty is shown in blue.
     
    \subsection*{\HeCsixty{4} INS}
\begin{figure}[hbt!]
          \centering
          \includegraphics[width=0.8\textwidth]{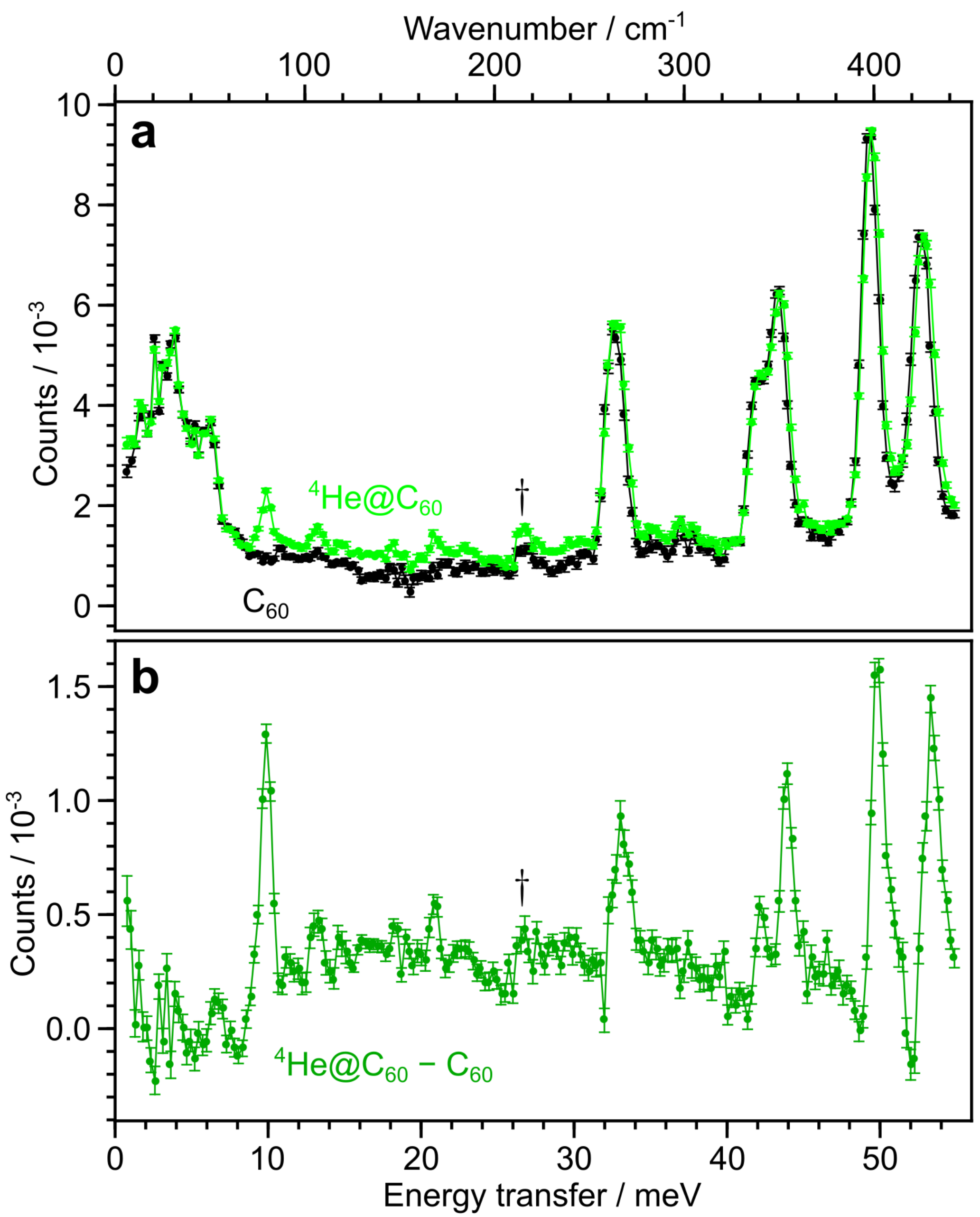}
          \caption{\textbf{(a)} IN1 LAGRANGE INS spectra of: 294 mg \HeCsixty{4} (f=40\%) in green and 293 mg \Csixty in black, at 2.7\kel. \textbf{(b)} in dark-green the difference \HeCsixty{4}$-$\Csixty, green$-$black, from (a) is shown.
          Counts are normalised with respect to the monitor neutron count. An experimental artefact is marked with a dagger ($\dag$).
          %
          }
          \label{fig:4HeC60_INS_IN1_fig_diffDKGRN_4HeC60GRN_C60x0.92BLK_FIG}
\end{figure}

     The low temperature INS spectra of 294 mg \HeCsixty{4} (f=40\%) and 293 mg \Csixty are shown in fig.~\ref{fig:4HeC60_INS_IN1_fig_diffDKGRN_4HeC60GRN_C60x0.92BLK_FIG} (a), in green and black respectively.

The \Csixty spectrum is scaled to best match the \HeCsixty{4} spectrum, in order to obtain optimum background subtraction. In fig.~\ref{fig:4HeC60_INS_IN1_fig_diffDKGRN_4HeC60GRN_C60x0.92BLK_FIG} (b) the difference between \HeCsixty{4} and \Csixty is shown in dark-green. An experimental artefact, present in both \HeCsixty{4} and \Csixty spectra, is marked with a dagger ($\dag$).

%

\section*{Quantum Theory}


    \subsection*{Quantum mechanics for spherically symmetric potentials}
    
    The time-independent \Schrodinger equation is given by~\cite{griffiths_introduction_2018}:
    
\begin{equation}\label{eq:SEq}
    \hat{H}(\vecr_1, \vecr_2\ldots) \psi_\vecq(\vecr_1, \vecr_2\ldots) = E_\vecq \psi_\vecq(\vecr_1, \vecr_2\ldots)
\end{equation}
    where the vectors $\vecr_1$, $\vecr_2\ldots$ are the coordinates of the particles, the quantum state is described by the eigenfunction $\psi_\vecq(\vecr_1, \vecr_2\ldots)$, $\vecq$ represents any set of quantum numbers ($\vecq=\{q_1,q_2\ldots\}$) and $E_\vecq$ is the eigenvalue (energy) of the eigenfunction $\psi_\vecq$. When applying the Hamiltonian operator $\hat{H}$ (eq.~\ref{eq:Ham1}) to the eigenfunction $\psi_\vecq$, one obtains the eigenfunction back and the eigenvalue $E_\vecq$ associated with it. The Hamiltonian operator $\hat{H}$ is given by:
    
\begin{equation}\label{eq:Ham1}
    \hat{H}(\vecr_1, \vecr_2\ldots) = -\sum_{i=1}^N \frac{\hat{p}_i^2}{2M_i} + V(\vecr_1, \vecr_2\ldots)
\end{equation}
where $\hat{p}_i$ and $M_i$ are the momentum operator and mass for particle $i$. In the absence of a potential $V$, eq.~\ref{eq:Ham1} is the Hamiltonian for a free particle.
For the cases studied here the potential is not zero, and for endofullerenes $V$ is the confining potential which keeps the endohedral moiety enclosed. The meaning of $V$ is such that a quantum particle $i$ at coordinates $\vecr_1$ has potential energy given by $V(\vecr_1)$.

 In the cases of noble gas endofullerenes studied here, there is only one confined particle so the eigenfunction depends only on three spatial variables, $\psi_\vecq(\vecr_1) = \psi_\vecq(r,\theta,\phi)$ in spherical polar coordinates. The variables are separated into a radial part and angular part:
    \begin{equation}\label{eq:wfRY}
    \psi_\vecq(r,\theta,\phi)=R_\vecq(r)Y_\vecq(\theta,\phi)
    \end{equation}
  The \Schrodinger equation is factorised into a radial (eq.~\ref{eq:radialEQ}) equation and an angular equation (eq.~\ref{eq:angularEQ})~\cite{griffiths_introduction_2018}, where M is the mass of the particle:
    \begin{align}
    \radialEQ   \label{eq:radialEQ}\\
     \; u_\vecq(r) & =rR_\vecq(r)\\
    \angularEQ    \label{eq:angularEQ}
    \end{align}
    The wavefunctions which solve the angular equation are spherical harmonics~\cite{griffiths_introduction_2018} and are given below: 
    \begin{align}
    \YsphericalHarmonics    \label{eq:Ysh}\\
    \PLegendreAssocFct   \label{eq:PLegendreAssocFct}\\
     \PLegendrePoly   \label{eq:PLegendre}
    \end{align}
    Here $P_\ell ^m (x)$ are the associated Legendre functions and $P_\ell(x)$ are the Legendre polynomials. Thus, for the angular part of the \Schrodinger equation the solutions are known. $\ell$ is the angular momentum quantum number and $m$ is azimuthal quantum number.
    
  The radial solutions of the \Schrodinger equation depend strongly on the potential $V(r)$.

    \subsection*{3D quantum harmonic oscillator (HO)}\label{SI:theory_3D_HO}
    The spherically symmetric case of eq.~\ref{eq:Ham1}, where $V(r)= V_2 r^2= \frac{1}{2}k r^2$, is well known as the 3D quantum Harmonic Oscillator (HO), where $k=M\omega_0^2$. 
     The Hamiltonian ($\hat{H}_0$) for the spherically symmetric 3D harmonic oscillator is:
         \begin{equation}
            \hat{H}_0 =-\frac{\hat{p}^2}{2M} + V_2 r^2 =-\frac{\hat{p}^2}{2M} +\frac{k}{2}r^2\label{eq:Ham3DHO}
         \end{equation}
         
         To find the wavefunctions of the 3D harmonic oscillator (HO), we have to separate them into radial and angular parts\cite{flugge_practical_1999,bransden_quantum_2000}:
         \begin{equation}
             \PSInlm(r,\theta,\phi)=\Rnl(r)\Ylm(\theta,\phi)\label{eq:wfHOtot}
         \end{equation}
         
         The angular part of the wavefunction are the spherical harmonics $\Ylm$ given in eq.~\ref{eq:Ysh}. The radial part of the wavefunction for the 3D HO is given below\cite{flugge_practical_1999,bransden_quantum_2000}: 
        \begin{align}
         R_{n\ell}(r) &= \mathcal{N}(n,\ell,\beta) \,e^{\frac{-\beta r^2}{2}}\, \left(\beta r^2\right)^\frac{\ell}{2} L^{\ell+\frac{1}{2}}_\frac{n-\ell}{2}\left[\beta r^2\right] \label{eq:Rwf3DHO}\\
        %
        %
          \mathcal{N}(n,\ell,\beta)  &= \sqrt{\frac{2\left(\frac{n-\ell}{2}\right)!}{\left(\frac{n+\ell+1}{2}\right)! }}\beta^\frac{3}{4}   \label{eq:Rwf3DHOAnorm }\\
          %
          %
          \beta & = \frac{M \omega_0}{\hbar}\\
          %
        %
        \omega_0 & =\sqrt{\frac{k}{M}}
        \end{align}
        where $\mathcal{N}(n,\ell,\beta)$ is a normalisation constant and $L^{\ell+\frac{1}{2}}_\frac{n-\ell}{2}[x]$ are the generalised Laguerre polynomials.

        The eigenvalues $E_{n \ell}$ of the 3D HO are given by:
        \begin{equation}
        \begin{split}
       E_{n \ell} & = \hbar \omega_0 (n + \frac{3}{2})=\hbar \omega_0 (\ell + 2 n_r + \frac{3}{2})\label{eq:eval3DHO}\\
       n & = \ell+2n_r\\
        \ell & = 0,1,2,3,...\; \text{and} \; n_r = 0,1,2,3,... \;  \; 
        \end{split}
        \end{equation}
Which means:
         \begin{align*}
             \ell & = 0, 2, ..., n-2, n; \; \textrm{if}\; n \textrm{ = even} \\
             \ell & = 1, 3, ..., n-2, n; \; \textrm{if}\; n \textrm{ = odd} \\
         \end{align*}
Since the potential is spherically symmetric, the $2\ell+1$ degeneracy of $m$ substates of each $\ell$ state is not broken. The usual $n$ quantum number of a harmonic oscillator is given by $n = \ell+2n_r$; where the quantum number $n_r$ represents the number of radial nodes that the wavefunction $\Rnl(r)$ posses.
Furthermore, eigenstates of different $\ell$ but same $n$ quantum numbers have accidental degeneracy because the potential is purely harmonic.

    \subsection*{3D polynomial oscillator}
    
    
The 3D HO quadratic potential ($V_2 r^2=\frac{k}{2}r^2$) is not sufficient to describe the confining potential of \HeCsixty{}. We use the more general polynomial form:
    \begin{equation}\label{eq:VrPOLYsum}
    V(r) = V_2 r^2 + V_4 r^4 + V_6 r^6
    \end{equation}
where $\{V_2,V_4,V_6\}$ are polynomial coefficients, assuming spherical symmetry. There are no known analytic solutions of the radial \Schrodinger equation (eq.~\ref{eq:radialEQ}) for polynomial potentials. Numerical linear algebra diagonalisation is used to calculate the solutions, which converge given a large enough basis set.
        
An element of the matrix representation of a given Hamiltonian $\hat{H}^{(a)}$ is defined as follows~\cite{griffiths_introduction_2018}:
\begin{equation}
   \matrixELEMENTgeneral\label{eq:matrixELMTgeneral}
\end{equation}
Eq.~\ref{eq:matrixELMTgeneral} is simplified by the orthonormality of the spherical harmonics (eq.~\ref{eq:Yorthonormal} below)~\cite{griffiths_introduction_2018}:
\begin{equation}
 \int_0^\pi\int_0^{2\pi} \Ylm^*\, Y_{\ell' m'} \, \sin(\theta) \; d\theta \; d\phi = \delta_{\ell \ell'} \, \delta_{m m'} \label{eq:Yorthonormal}
\end{equation}
Eq.~\ref{eq:Yorthonormal} dictates that the final Hamiltonian is block diagonal, with each different $\ell$ subspace making up the blocks.

\begin{figure}[hbt!]
          \centering
          \includegraphics[width=1\textwidth]{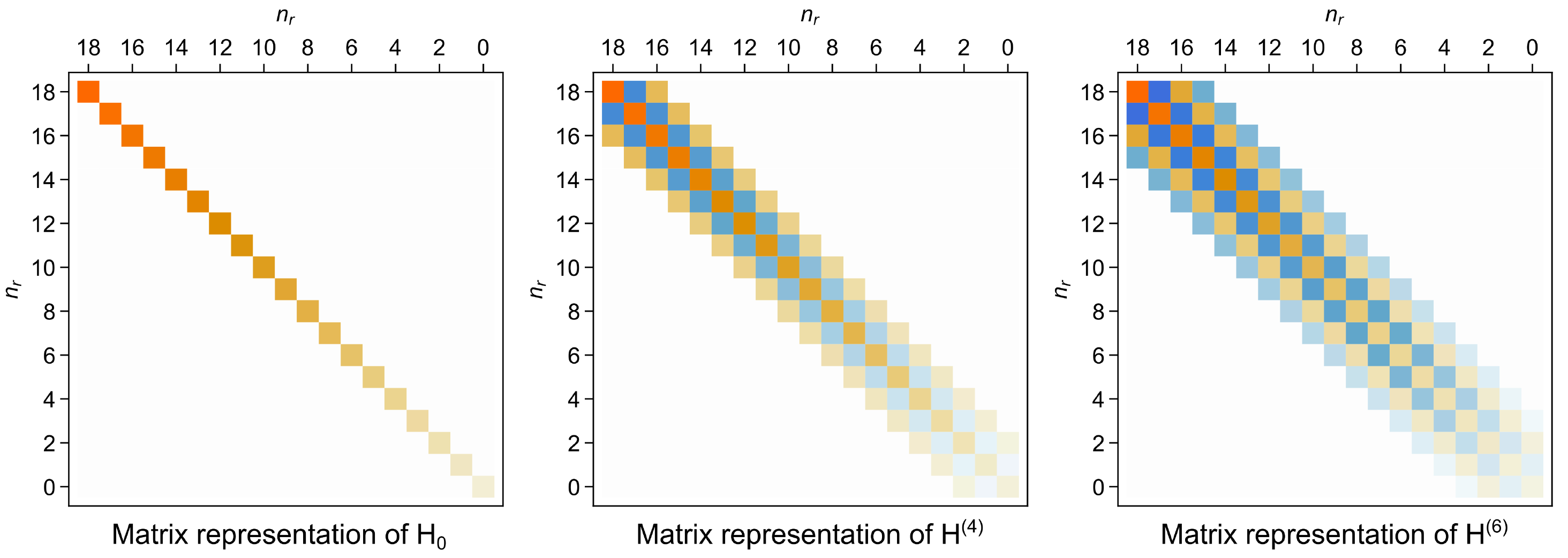}
          \caption{ Matrix plot representations for $\hat{H}_0$, $\hat{H}^{(4)}$, $\hat{H}^{(6)}$ from eq.~\ref{eq:INS_THz_introQM_H_H0a4a6}, with $n_r=0,1,\ldots18$ and $\ell = 0$. Similar plots are obtained for the other $\ell$ values. The color intensity scales with the magnitude of the matrix element. Orange represents positive matrix elements and blue negative.
          }
          \label{fig:INS}
\end{figure}

When using the 3D HO as a basis, with the wavefunctions from eq.~\ref{eq:Rwf3DHO} as basis states, the matrix representation of $\hat{H}_0$ (eq.~\ref{eq:Ham3DHO}) is diagonal:
\begin{equation}
        \braket{\PSInlm|\hat{H}_0|\psi_{n' \ell'm'}}=E_{n \ell} \, \delta_{n n'} \, \delta_{\ell \ell'} \, \delta_{m m'} \label{eq:eval3DHOdiagonal}
\end{equation}
where $E_{n \ell}$ are the 3D HO eigenvalues (eq.~\ref{eq:eval3DHO}) and the Kronecker delta $\delta_{ab}$ = 1 if $a=b$ or = $0$ if $a\neq b$. See fig.~\ref{fig:INS} for the form of the matrix representation of $\hat{H}_0$.
        
The left-over terms in the potential, $ V_4 r^4 + V_6 r^6$ (eq.~\ref{eq:VrPOLYsum}), are written as:
\begin{equation}
            \hat{H}^{(4)}= V_4 r^4\; \mathrm{and} \; \hat{H}^{(6)}=V_6 r^6
\end{equation}
By using eq.~\ref{eq:matrixELMTgeneral} all matrix elements of $\hat{H}^{(4)}$ and $\hat{H}^{(6)}$ are computed analytically, using the Wolfram Mathematica software. Matrix representations of $\hat{H}^{(4)}$ and $\hat{H}^{(6)}$ (in the 3D HO basis) are seen in fig.~\ref{fig:INS}, for $\ell=0$ and $n_r=0,1,...,18$.
       
%
The Hamiltonian for the 3D polynomial oscillator discussed here is $\hat{H}$:
\begin{equation}
            \hat{H}= \hat{H}_0+\hat{H}^{(4)}+\hat{H}^{(6)}\label{eq:INS_THz_introQM_H_H0a4a6}
\end{equation}
        
        
        Performing numerical diagonalisation on the matrix representation of $\hat{H}$ gives the final eigenvalues and eigenfunctions which best describe the Helium atom confined inside \Csixty. Within spherical symmetry ($\hat{H}$ does not depend on $\theta$ and $\phi$) each energy level of the 3D polynomial oscillator Hamiltonian (eq.~\ref{eq:INS_THz_introQM_H_H0a4a6}) is $2\ell+1$ times degenerate. Therefore, it is sufficient to only include the $\psi_{n\ell 0}(r,\theta,\phi)$ wavefunctions in the basis set. 


        

\subsection*{THz oscillator line strengths}


To interpret the intensities of the THz peaks we need to make use of the Fermi's golden rule~\cite{griffiths_introduction_2018} 
and assume some type of induced dipole moment operator since the He atom does not have a permanent dipole moment. 
The dipole moment associated with the translational motion of the He atom inside the \Csixty cage is assumed to be linear in the displacement from the cage centre $\vecr$ for simplicity. Written in spherical harmonics it reads:
\begin{equation}
d_{1m}  =\sqrt{\frac{4\pi}{3}} A_{1m} \, r \, Y_{1m}(\theta,\phi), \label{eq:THzdipMOM}
\end{equation}
where $A_{1m}$ is the coefficient for the dipole moment operator and $m  =-1, 0, +1$. 
The \HeCsixty{} cages are randomly orientated in the powder, so  the dipoles are randomly oriented relative to the electric field of radiation.
Averaging of $(\mathbf{E}\cdot \mathbf{d})$ over all orientations  gives $E^2 (d_x^2 + d_y^2+ d_z^2)/3$~\cite{loudon_quantum_2000}. 
Since spherical symmetry is assumed,  the induced dipole moment does not depend on the direction of displacement $\vecr$  and $A_{1m}$ is independent of $m$, $A_{1m}\equiv A_1$. 
Thus, all three components of the induced dipole moment are the same.


The THz absorption line area is written using the Fermi's golden rule~\cite{griffiths_introduction_2018,mamone_rotor_2009,ge_interaction_2011}, for  light polarised  linearly in the $z$ direction ($d_z\equiv d_{10}$):
    \begin{equation}
     \int_{\omega_{fi}} \alpha_{fi}(\omega) \mathrm{d} \omega 
 = N f \frac{2\pi^2  }{  h \epsilon_0 c_0 \eta  }
 \left({\frac{\eta^2 +2  }{ 3  }}\right)^2 \omega_{fi} \left(p_i-p_f\right) \sum_{m_i, m_f} \left|\bra{n_{f}, l_f, m_f}d_{10}\ket{n_{i},l_i, m_i}\right|^2, \label{eq:THz_abs_linear}  \end{equation}
where each $\ell$ state is ($2 \ell + 1$) is degenerate in $m$, so summation is done over all $m_i$ and $m_f$. 
 The integral is taken over the whole frequency range $\omega_{fi} $ spanning the THz transition $\ket{i}\rightarrow \ket{f}$. 
$N=1.48 \times 10^{27} \mathrm{m}^{-3}$ is the number density of molecules in solid \Csixty, $f$ is the filling factor of the endofullerene, $c_0$ is the speed of light in vacuum, $\epsilon_0$ is the permittivity of vacuum, $\eta$ is the index of refraction (for \Csixty $\eta = 2$, see  ref.~\citenum{homes_effect_1995} and ref.~47 in ref.~\citenum{shugai_infrared_2021}). 
Since in spherical symmetry $d_x=d_y=d_z$,   eq.~(\ref{eq:THz_abs_linear}) is valid for  randomly polarised radiation as well.

The factor $(\eta^2 +2)/3$ is the enhancement of electric field felt by the oscillator  in a dielectric medium~\cite{dexter_absorption_1956}. 
$\omega_{fi}=(E_f-E_i)/h c_0$, where $E_i$ and $E_f$ are the eigenvalues of the initial and final state and $h$ is the Planck constant.


$p_i$ and $p_f$ are the thermal Boltzmann populations of the initial and final states:

 \begin{equation}\label{eq:Boltzmann_with_degen}
      p_i = \frac{ e^{-E_i/k_B T}}{\sum_j (2 \ell_j +1)\,e^{-E_j/k_B T}}.
 \end{equation}   
where $k_B$ is the Boltzmann constant and $T$ is the temperature.

The  sum over $m$ in Eq.~(\ref{eq:THz_abs_linear}) equals to~\cite{gordy_microwave_1984}: 
   \begin{equation}
\sum_{m_i, m_f}\left|\bra{n_{f}, \ell_f, m_f}d_{10}\ket{n_{i},\ell_i, m_i}\right|^2= \frac{1}{3}\left|\bra{n_{f}, \ell_f}| d_{1} |\ket{n_{i}, \ell_i}\right|^2, \label{eq:M_elem_reduced}  
\end{equation}
where the dependence on  $m$ has disappeared and $\bra{n_{f}, \ell_f} | d_{1} | \ket{n_{i}, \ell_i} $ 
is the reduced matrix element of $d_{1q}$~\cite{zare_angular_1988}.




The angular part of the reduced matrix element is~\cite{mamone_experimental_2016,zare_angular_1988}:
\begin{equation}\label{eq:matrix_elem_reduced}
    \bra{\ell_f}|T_k|\ket{\ell_i}= (-1)^{-\ell_f} \sqrt{\frac{(2\ell_f+1)(2k+1)(2\ell_i+1)}{4\pi}} \left( \begin{array}{ccc}\ell_f & k & \ell_i \\0 & 0 & 0\end{array}\right),
\end{equation}
where $T_k$ is a spherical tensor operator of rank $k$ and the six symbols in the brackets denote Wigner $3j$-symbol~\cite{zare_angular_1988}. The
$3j$-symbol is zero if $|\ell_f-\ell_i|\leq k \leq \ell_f+\ell_i$ is not satisfied.
Another property of the $3j$-symbol with  $ m_f = q = m_i = 0 $ is that  it is non-zero only if $\ell_f+ k +\ell_i$ is even. 
For the dipole moment $k=1$, ergo, the selection rule $\Delta \ell = \pm1$ for THz absorption arises.

\subsection*{Parameter fitting}
        
The experimental THz absorption spectrum was fitted using Gaussian line shapes to find the line areas, line widths and frequencies.
A synthetic experimental spectrum $y(\omega_{n})$,  the distance between the points in the spectrum  $\omega_n - \omega_{n-1}=\Delta\omega / 4$, was then generated consisting of lines with equal linewidths, $\Delta \omega = 1$\wn, while keeping the line areas and frequencies of the original experimental lines.
The synthetic spectrum approach was needed as our model did not include any line broadening mechanism.

The following  lines of the synthetic spectrum, format (frequency [\wn], area [cm$^{-2}$]), were used for the parameter fitting.

\HeCsixty{3} at 5~K: (96.77, 10.72),

\HeCsixty{3} at 125~K:  (96.88, 2.61), (105.71, 3.30), (113.62, 2.25), (119.76, 0.56), (121.63, 1.79), (128.13, 0.32), (137.73, 0.52),

\HeCsixty{4} at 5~K: (81.27, 7.86),

\HeCsixty{4} at 100~K: (81.32, 2.082),(88.48, 2.40),(94.795, 1.46),(100.84, 1.63), (105.89, 0.488), (110.38, 0.11), (113.95, 0.42),(125.73, 0.34).

For a given model and basis, matrix elements of the Hamiltonian (eq.~\ref{eq:INS_THz_introQM_H_H0a4a6}) 
 and the dipole operator (eq.~\ref{eq:THzdipMOM}) were evaluated analytically in a symbolic form using Mathematica software.  
At each step of minimizing chi squared, $\chi^2=\sum [y-f(\omega_{n},\{\kappa\}) ]^2 $, the numerical values were substituted for symbols and the Hamiltonian diagonalized numerically.
Here $f(\omega_{n},\{\kappa\})$ is the theoretical spectrum  with same linewidth and lineshape as the synthetic experimental spectrum;
 $\{\kappa\}$  is  the set of fit parameters:  Hamiltonian and dipole operator parameters, $ \{ V_2, V_4, V_6, A_{10} \}$.
As first partial derivatives are zero at the best fit, we used second derivatives to calculate the error margins $\Delta \kappa_i $ of the fit parameters:
\begin{equation}
\Delta \kappa_i = \sqrt{2\chi^2\left(\frac{\partial^2 \chi^2}{\partial \kappa^2_i}\right)^{-1}}.
\end{equation}

In the spherical approximation the energy does not depend on  $m$.
Therefore it is practical to use a reduced basis and reduced matrix elements of spherical tensor operator $T_{kq}$ of rank $k$ which are independent of $m $ and $q$~\cite{gordy_microwave_1984}. 
The rank of potential spherical operators is $k=0$ and the rank of dipole operator is $k=1$.
In the reduced basis  the number of states is smaller by  factor $2\ell+1$ for each $\ell$. 
In such reduced basis   there are  100    states for $n_\mathrm{max}=18$. 

Forcing  $V_6=0$ increased the $\chi^2$, as compared  to the  fit with all three potential parameters, three times and two times  for \Helium{3} and \Helium{4}, respectively.

The calculated energy levels for the best-fit parameters (with $n_\mathrm{max}=18$) are given in table~\ref{tab:3He_energies} and ~\ref{tab:4He_energies}. 

\newpage

\begin{table}[hbt!]
	\centering
	\caption {
		\label{tab:3He_energies}Translational energy levels of   \HeCsixty{3} obtained from the fit of 125\,K THz absorption spectrum.
		Translational energy $E$, the angular momentum quantum number $\ell$,  and the amplitude squared  $|\xi|^2$ of the main component of eigenstate with the principal quantum number $n$. The energies are given relative to the ground state. The potential parameters are given in Table 1 of the main text. 
	}
	\setlength{\tabcolsep}{15pt}
	\begin{tabular}{lccc}
		
		$E/\wn$ & $\ell$ & $n$  & $|\xi|^2$ \\ \hline
		0   &  0 & 0&  0.95\\
		96.9& 1 & 1   & 0.88\\
		202.6& 2&2  & 0.78\\
		218.2& 0 & 2  & 0.66\\
		316.3 & 3 &3	 &  0.67\\
		340.6 & 1 &	3 &  0.45\\
		437.3 & 4 & 4 & 0.55\\
		469.6 & 2 & 6 &  0.4 \\
		483.2 & 0 &6 &  0.4 \\
		565.2 & 5 & 5 &  0.44 \\
		604.8 & 3 & 7 &0.36 \\
		626.4 & 1 & 7 &  0.32 \\ 
	\end{tabular}
\end{table}

\begin{table}[hbt!]
	\centering
	\caption{\label{tab:4He_energies} Translational energy levels of \HeCsixty{4} obtained from the fit of 100\,K THz absorption spectrum. 
		Translational energy $E$, the angular momentum quantum number $\ell$,  and the amplitude squared  $|\xi|^2$ of the main component of eigenstate with the principal quantum number $n$.
		The energies are given relative to the ground state. The potential parameters are given in Table 1 of the main text.
		\\
	}
	\setlength{\tabcolsep}{15pt}
	\begin{tabular}{lccc}
		
		$E/\wn$ & $\ell$ & $n$  & $|\xi|^2$ \\ \hline
		0   &  0 & 0&  0.95\\ 
		81.4 & 1 & 1   & 0.89\\ 
		169.8& 2 &	2  & 0.8\\
		182.1 & 0 & 2  & 0.69\\ 
		264.5 & 3 &	3 &  0.69\\
		283.8 & 1 &	3 &  0.49\\ 
		365.2 & 4 & 4 & 0.58\\
		390.6 & 2 & 6 &  0.4 \\
		401.4 & 0 & 6 &  0.41 \\ 
		471.3 & 5 & 5 &  0.48 \\
		502.5& 3 &7 & 0.38 \\
		519.6 & 1 & 7 &  0.35 \\ 
		582.5 & 6 &6 & 0.38\\
		618.8 & 4 & 8 & 0.31\\
		641.7 & 2 & 10 & 0.31\\
		651.4 & 0& 10 & 0.31\\
	\end{tabular}		
\end{table}

\section*{Empirical potentials}\label{SI:section_empirical_potentials}

The confining potentials in this section are obtained by summing 60 two-body potentials between the enclosed Helium and the Carbon atoms constituting the \Csixty cage.

To obtain the confining potential $V(r)$, first a 3D structure of \Csixty is generated with a given radius and HP \& HH bond lengths. A two-body interaction potential $U(\rho)$ between the carbon atoms of the cage and the endohedral Helium atom is chosen. The confining potential at any one point is obtained by summing all the 60 He$\,\cdots\,$C two-body interactions.  To get the confining potential along an axis, the position of the Helium atom is moved along that axis and then the potential is computed for each position.


The structural parameters of the cage were taken from neutron diffraction measurements on \Csixty, from ref.~\citenum{leclercq_precise_1993}. The C-C bond lengths~\cite{leclercq_precise_1993} used in calculating the \HeCsixty{} confining potential are: $p= 1.4597 \pm 0.0018$ \AA{} (hexagon-pentagon edge) and $h= 1.3814 \pm 0.0027$ \AA{} (hexagon-hexagon edge); which gives a radius of the \Csixty $ R = 3.547 \pm 0.005$ \AA{}.


    
    
\subsection*{Lennard-Jones 6-12 potential (ref.~\citenum{carlos_interaction_1980})}
        
        The empirical He$\,\cdots\,$C two-body Lennard-Jones (6-12) interaction potential, $U^{LJ}_{6-12}(\rho)$, obtained from He/graphite scattering experiments (ref.~\citenum{carlos_interaction_1980}), is defined in eq.~\ref{eq_LJ_6_12} below, with parameters given in table~\ref{table_LJ_6-12_parameters}.
    	\begin{equation}
    	U^{LJ}_{6-12}(\rho)=4\epsilon\left[\left(\frac{\sigma}{\rho}\right)^{12}-\left(\frac{\sigma}{\rho}\right)^{6}\right]\label{eq_LJ_6_12}
    	\end{equation}
where $\rho$ is the interatomic distance between Helium and Carbon, $\epsilon$ is the potential well depth, and $\sigma$ is the interatomic distance at which the potential energy is zero.
    	
\begin{table}[h!]
    	\centering \caption{Parameters for the empirical Lennard-Jones (6-12) two-body potential ($U^{LJ}_{6-12}(\rho)$ from eq.~\ref{eq_LJ_6_12}) between carbon and Helium, from He/graphite scattering (ref.~\citenum{carlos_interaction_1980}).
    	\\
    	}
    	\label{table_LJ_6-12_parameters}
    	\begin{tabular}{lcc}
    		
    		Carbon-Atom & $ \sigma$ (\AA{}) & $ \epsilon$ (meV) \\
    		\hline
    		Helium & 2.74 & 1.40\\
    		
    		\end{tabular}
\end{table}

\subsection*{Lennard-Jones 6-8-12 potential (ref.~\citenum{carlos_interaction_1980})}
The empirical He$\,\cdots\,$C two-body Lennard-Jones (6-8-12) interaction potential, $U^{LJ}_{6-8-12}(\rho)$, obtained from He/graphite scattering experiments (ref. ~\citenum{carlos_interaction_1980}), is defined in eq.~\ref{eq_LJ_6_8_12} below, with parameters given in table~\ref{table_LJ_6-8-12_parameters}.

    	\begin{equation}
    	U^{LJ}_{6-8-12}(\rho)=\frac{\epsilon}{2s+3}\left[\left(4s+3\right)\left(\frac{\rho_m}{\rho}\right)^{12}-6s\left(\frac{\rho_m}{\rho}\right)^{8}-6\left(\frac{\rho_m}{\rho}\right)^{6}\right] \label{eq_LJ_6_8_12}
    	\end{equation}
where $\rho$ is the interatomic distance between Helium and Carbon. The equilibrium position (minimum energy) is denoted by $\rho_m$. The parameter $s$ specifies the ratio of the two attractive terms (6 and 8) at $\rho_m$.
        
        \begin{table}[h!]
    	\centering \caption{Parameters for the empirical Lennard-Jones (6-8-12) two-body potential ($U^{LJ}_{6-8-12}(\rho)$ from eq.~\ref{eq_LJ_6_8_12}) between carbon and Helium, from He/graphite scattering (ref.~\citenum{carlos_interaction_1980}).
    	\\
    	}
    	\label{table_LJ_6-8-12_parameters}
    	\begin{tabular}{lccc}
    	
    		Carbon-Atom & $\rho_m$ (\AA{}) & s & $ \epsilon$ (meV) \\
    		\hline		
    		Helium & 3.68 & 2.45 & 1.12\\
    		
    		\end{tabular}
    	\end{table}

\subsection*{Lennard-Jones 6-12 potential (ref.~\citenum{pang_endohedral_1993})}

        The He$\,\cdots\,$C two-body Lennard-Jones interaction potential ($U^{LJ}(\rho)$), from ref.~\citenum{pang_endohedral_1993}, is defined in eq.~\ref{eq_LJ_pot_1993} below, with parameters given in table~\ref{table_LJ_parameters}.

    	\begin{equation}
    	U^{LJ}(\rho)=4\epsilon\left[\left(\frac{\sigma}{\rho}\right)^{12}-\left(\frac{\sigma}{\rho}\right)^{6}\right]\label{eq_LJ_pot_1993}
    	\end{equation}
where $\rho$ is the interatomic distance between Helium and Carbon. $\epsilon$ is the potential well depth and $\sigma$ is the interatomic distance at which the potential energy is zero.
    	
    	\begin{table}[h!]
    	\centering \caption{Parameters for the Lennard-Jones two-body potential ($U^{LJ}(\rho)$ from eq. \ref{eq_LJ_pot_1993}) between carbon and helium, from ref.~\citenum{pang_endohedral_1993}.\\ }
    	\label{table_LJ_parameters}
    	\begin{tabular}{lcc}
    		
    		Carbon-Atom & $ \sigma$ (\AA{}) & $ \epsilon$ (kJ/mol) \\
    		\hline		
    		Helium & 2.971 & 0.1554   \\
    		
    		\end{tabular}
    	\end{table}	
    	
            
\subsection*{Modified Buckingham potential (ref.~\citenum{jimenezvazquez_equilibrium_1996})}
        
        Another two-body He$\,\cdots\,$C interaction potential is the modified-Buckingham (mB) potential, $U^{mB}(\rho)$, from ref.~\citenum{jimenezvazquez_equilibrium_1996}, which is obtained from the MM3 molecular mechanics program~\cite{lii_molecular_1989}. It is defined in eq.~\ref{eq_mod_Buckingham_pot}, with parameters given in table~\ref{table_Buckingham_potential_MM3_parameters}.
    
    	\begin{align}
    	U^{mB}(\rho)=&\epsilon'\left[2.9*10^5*exp\left(\frac{-12.5}{a}\right)-2.25*a^6\right]\; (\textrm{if}\;a\leq 3.311)\label{eq_mod_Buckingham_pot}\\ 
    					= & 336.176*\epsilon'*a^2\;(\textrm{if}\;a\geq 3.311)
    	\end{align}
    	where $\rho$ is the interatomic distance between Helium and Carbon; $a=\frac{r_i+r_j}{\rho}$ and $\epsilon'=\sqrt{\epsilon_i\epsilon_j}$ and the parameters for the modified-Buckingham potential are found in table \ref{table_Buckingham_potential_MM3_parameters}.
    	
    	\begin{table}[h!]

    		\centering \caption{MM3 modified-Buckingham He$\,\cdots\,$C two-body potential ($U^{mB}(\rho)$ from eq. \ref{eq_mod_Buckingham_pot}) parameters, from ref.~\citenum{jimenezvazquez_equilibrium_1996}.\\}
    		\label{table_Buckingham_potential_MM3_parameters}
    		\begin{tabular}{lcc}
    		
    			Atom & $r_i$ (\AA{}) & $\epsilon_i$ (kJ/mol) \\
    			\hline		
    			C & 1.96 & 0.234   \\
    		
    			He & 1.53 & 0.109  \\
    			
    		\end{tabular}
    	\end{table}

        
    \subsection*{Comparison of two-body interaction potentials}
        
\begin{figure}[hbt!]
          \centering
          \includegraphics[width=1\textwidth]{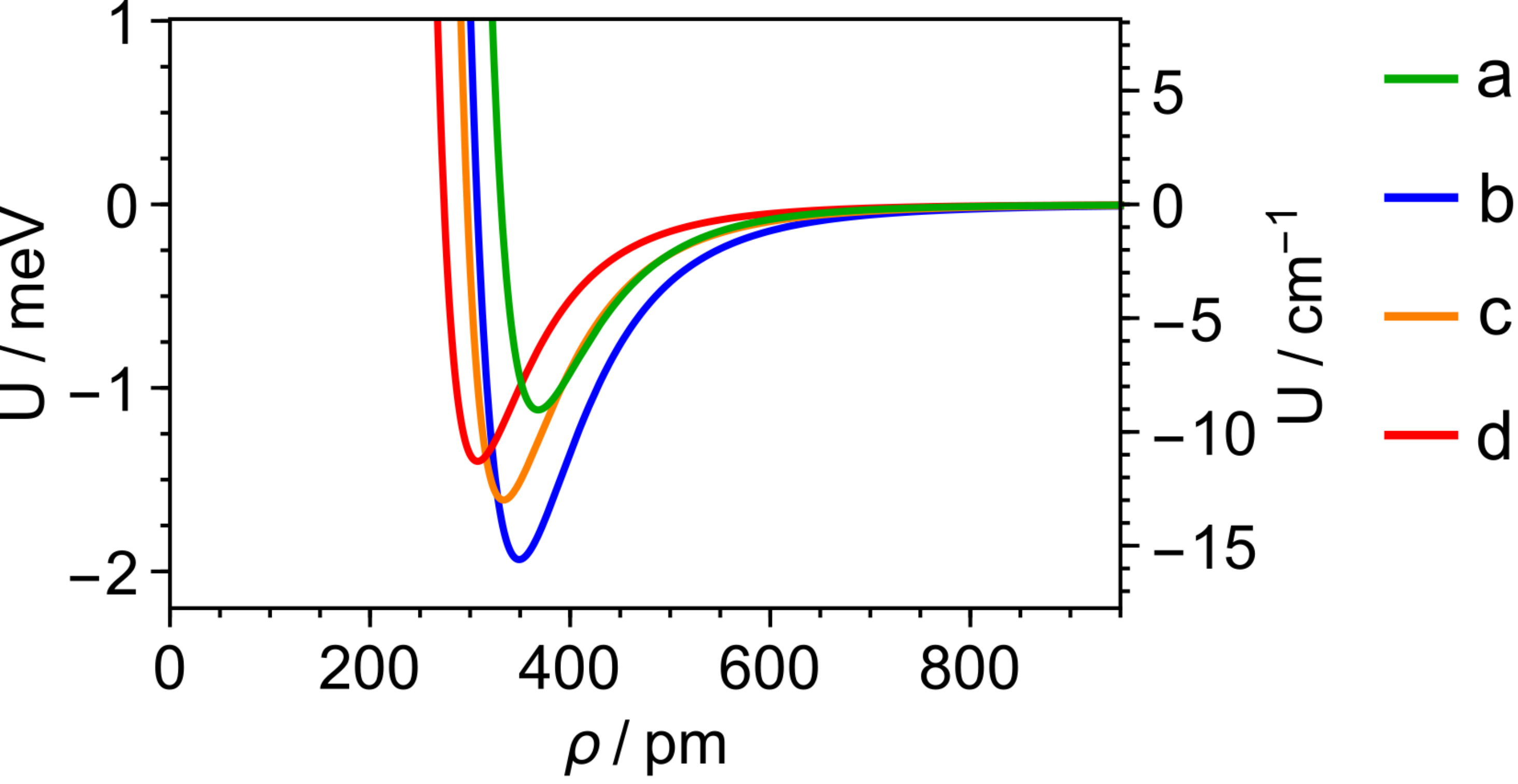}
          \caption{
          Helium$\,\cdots\,$Carbon two-body interaction potentials $U(\rho)$ against interatomic distance $\rho$. 
          \textbf{(a, green)} Lennard-Jones 6-8-12 potential with parameters from reference~\citenum{carlos_interaction_1980}; \textbf{(b, blue)} Modified Buckingham potential (as implemented in the MM3 program~\cite{jimenezvazquez_equilibrium_1996,lii_molecular_1989}); \textbf{(c, orange)} Lennard-Jones 6-12 potential with parameters from reference~\citenum{pang_endohedral_1993}; \textbf{(d, red)} Lennard-Jones 6-12 with parameters from reference~\citenum{carlos_interaction_1980}.
        The potentials used in (a) and (d) have been used for the fitting of He/graphite scattering data~\cite{carlos_interaction_1980}.}
          \label{fig:Two-Body_potentials_HeC_Semi_and_Empirical_all}
        \end{figure}

        All the two-body He$\,\cdots\,$C interaction potentials $U(\rho)$ used in this study are plotted in fig.~\ref{fig:Two-Body_potentials_HeC_Semi_and_Empirical_all}.

\subsection*{Anisotropy of confining potentials}
\begin{figure}[hbt!]
          \centering
          \includegraphics[width=1\textwidth]{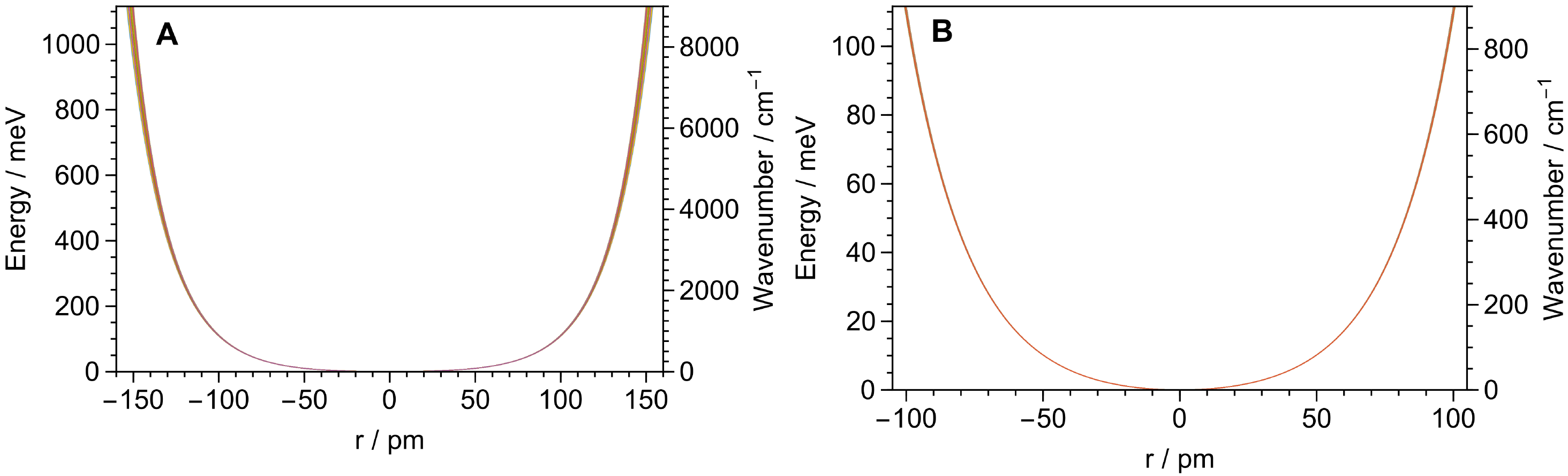}
          \caption{
          \HeCsixty{} confining potentials $V(r)$; obtained by summing the 60 He$\,\cdots\,$C empirical Lennard-Jones two-body interaction potentials $U(\rho)$ (eq.~\ref{eq_LJ_pot_1993} and ref.~\citenum{pang_endohedral_1993}), $r$ is the distance from the centre of the cage to the Helium atom. Multiple calculated potentials are shown, for moving the He along different directions: towards a C atom, towards the centre of a HH bond, towards the centre of a HP bond, towards the centre of pentagons/hexagons, etc. \textbf{(A)}: extended energy range showing small anisotropy of $V(r)$. \textbf{(B)}: restricted energy range probed in our experiments showing negligible anisotropy of $V(r)$. 
         }
          \label{fig:Rotation_C60_SemiEmpirical_LJ_Pang93_Two-Body_pot}
        \end{figure}

    All the \HeCsixty{} confining potentials $V(r)$ computed by summing the 60 He$\,\cdots\,$C two-body interaction potentials $U(\rho)$ are slightly anisotropic, since \Csixty is not a perfect sphere. However, in all cases the anisotropy was negligible in the energy range probed by our measurements; changing the direction along which the He is moved makes a very small difference to the confining potential. Figure~\ref{fig:Rotation_C60_SemiEmpirical_LJ_Pang93_Two-Body_pot} shows an example for the Lennard-Jones potential in eq.~\ref{eq_LJ_pot_1993} and ref.~\citenum{pang_endohedral_1993}.  Similar plots are obtained for all the other potentials. 
    
    

\section*{Computational chemistry}

We calculated the \HeCsixty{} radial potential $V(r)$ using Psi4.~\cite{psi4} Psi4 uses density-fitting (DF) to achieve favorable scaling with respect to system size by casting expensive electron repulsion integrals into linearly scaling auxiliary basis. 
Coupled-cluster theory~\cite{jensen_introduction_2017,HelgakerJorgensenOlsen} [CCSD, CCSD(T)] is generally considered to be \textit{state-of-the-art} method for interaction energy calculation but is practically not applicable for systems of this size. We calculated the potential with density functional theory (DFT) and M{\o}ller-Plesset perturbation (MP2) theory.~\cite{jensen_introduction_2017}
The DF procedure was used for both the self-consistent field (SCF) reference energy and the MP2 energy calculation.
By default, MP2 calculates same-spin and opposite-spin contributions to the correlation energy with different accuracy. The accuracy of the calculated energy was improved by applying the empirical spin-component-scaling factors (SCS).~\cite{SCS}

The DFT potential was calculated with
 the $\omega$B97X-V,~\cite{wB97XV,mardirossian2017thirty} and 
B3LYP,~\cite{lee_development_1988,vosko_accurate_1980,becke_densityfunctional_1993,stephens_ab_1994}
hybrid functionals. 
The first
functional is designed to handle non-covalent interactions with in-build contribution from the non-local VV10 correlation functional.~\cite{mardirossian2017thirty} The latter
functional is one of the most popular semi-empirical
hybrid functional.~\cite{mardirossian2017thirty} 
B3LYP 
cannot by default handle dispersion interaction and the Grimme D3 empirical dispersion correction with Beck-Johnson damping~\cite{grimme_consistent_2010,grimme_effect_2011} was applied to the functional.

\begin{figure}[hbt!]
  \centering
  \includegraphics[width=0.5\textwidth]{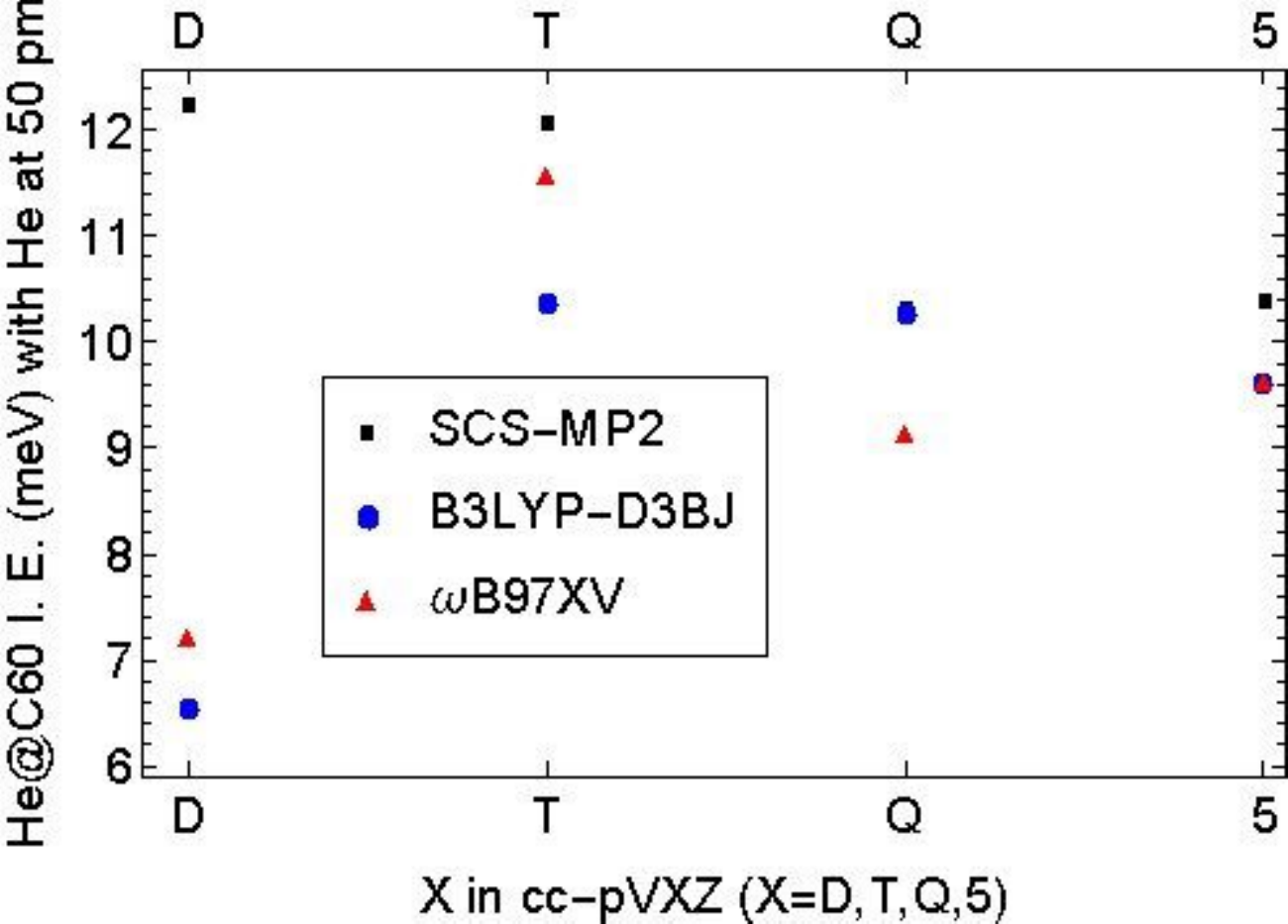}
  \caption{Quantum chemically calculated potentials (with respect to the minimum) as a function of the basis set size with Helium at $50$~pm from the center of the cage.}
  \label{fig:QchemBasis1}
\end{figure}

\begin{figure}[hbt!]
  \centering
  \includegraphics[width=0.5\textwidth]{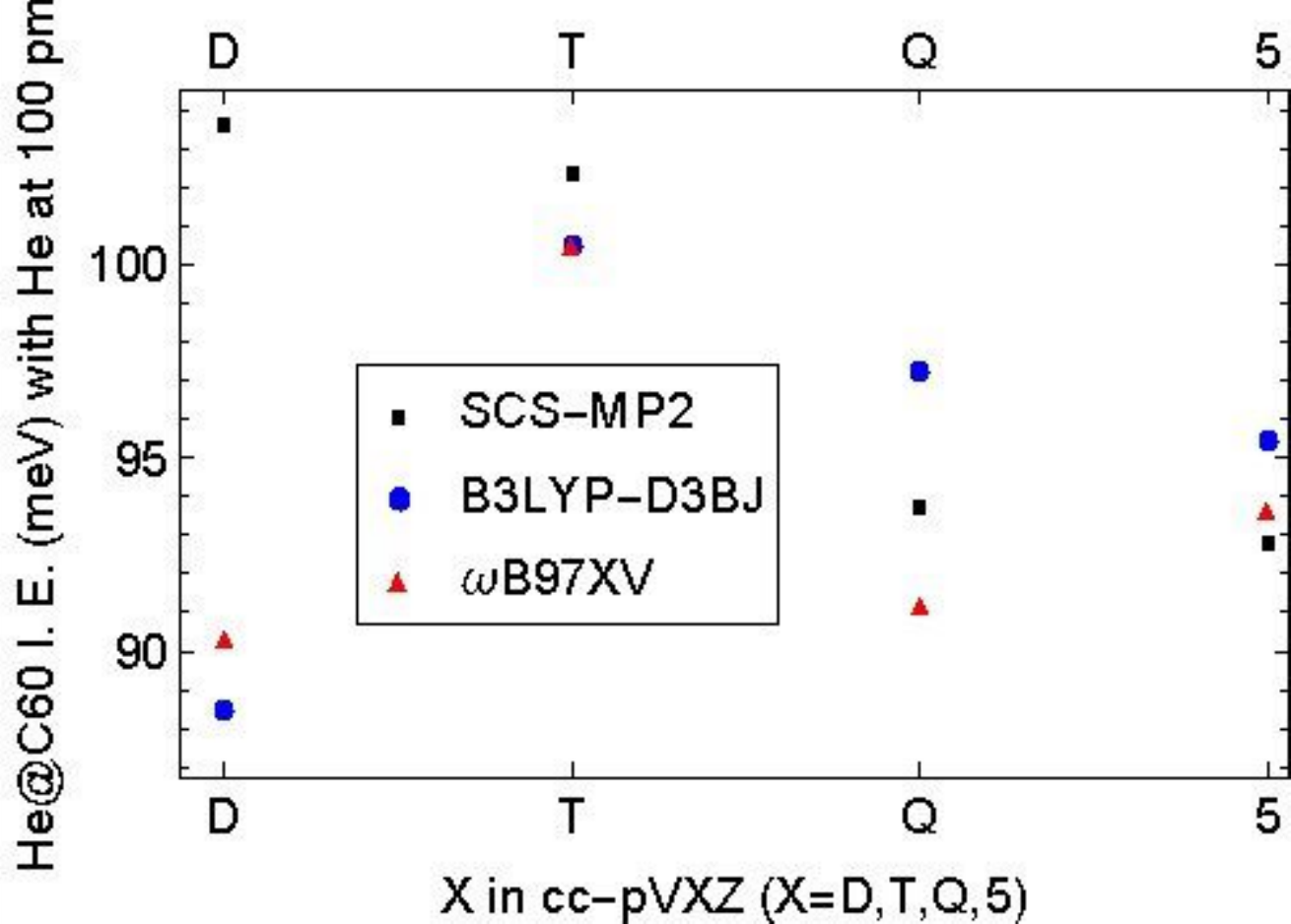}
  \caption{Quantum chemically calculated potentials (with respect to the minimum) as a function of the basis set size with Helium at $100$~pm from the center of the cage.}
  \label{fig:QchemBasis2}
\end{figure}

The potential was calculated as a function of the Helium distance from the centre of the cage, with \Csixty structural parameters from neutron diffraction measurements~\cite{leclercq_precise_1993} (see the main paper or section~\ref{SI:section_empirical_potentials} above). The Helium was moved along the axis between two inversion-related carbons.
The basis set convergence was tested with the Helium at $50$~pm and $100$~pm from the center of the cage. 
The calculations converged to a good approximation at X$=5$ with the correlation-consistent cc-pVXZ (X$=$D, T, Q, $5$, $6$) basis sets.~\cite{dunning1989gaussian,woon1994gaussian}  Figures~\ref{fig:QchemBasis1} and~\ref{fig:QchemBasis2} show the calculated potentials as a function of basis set size when the Helium atom is at $50$~pm and $100$~pm from the center of the cage respectively. The auxiliarity basis sets used by the DF algorithm were the defaults chosen by Psi4.
The counterpoise basis-set-superposition-error correction was applied in all the calculations,~\cite{van1994state} and all calculations were run in C$1$ symmetry. 
The dependence with respect to the direction the Helium is moved was investigated with MP2. The potential was calculated with the Helium moved along the three different symmetry axes of \Csixty and an axis between two inversion-related carbons. The difference between the calculated potentials was found to be negligible, with the Helium in the range of $0$~pm to $100$~pm from the center of the cage.
The potentials shown in the main paper were calculated in cc-pVQZ level with the Helium moved along the axis between two inversion-related carbons.

\bibliographystyle{rsc}
\bibliography{references/GRB-HeC60_arx, references/QChem-revised_arx}